\documentclass[english,showpacs,prl,aps,superscriptaddress,preprintnumbers,nofootinbib]{revtex4-1}
\usepackage[unicode=true,pdfusetitle,
 bookmarks=true,bookmarksnumbered=false,bookmarksopen=false,
 breaklinks=false,pdfborder={0 0 1},backref=false,colorlinks=false]
 {hyperref}
\usepackage[T1]{fontenc}
\usepackage[latin9]{inputenc}
\usepackage{babel}
\usepackage{amsmath}
\usepackage{amssymb}
\usepackage{graphicx}

\makeatletter

\providecommand{\tabularnewline}{\\}

\usepackage{epsfig}
\usepackage{amsfonts}

\makeatother

\begin{document}

\title{$J/\mathbf{\psi}$ production in hadron scattering: three-pomeron
contribution}

\author{E.~ Levin}
\email{leving@tauex.tau.ac.il, eugeny.levin@usm.cl}

\affiliation{Department of Particle Physics, School of Physics and Astronomy,
Raymond and Beverly Sackler Faculty of Exact Science, Tel Aviv University,
Tel Aviv, 69978, Israel}

\affiliation{Departemento de Física, Universidad Técnica Federico Santa María,
and Centro Científico-~~\\
 Tecnológico de Valparaíso, Avda. España 1680, Casilla 110-V, Valparaíso,
Chile}

\author{M. Siddikov}
\email{marat.siddikov@usm.cl}

\affiliation{Departemento de Física, Universidad Técnica Federico Santa María,
and Centro Científico-~~\\
 Tecnológico de Valparaíso, Avda. España 1680, Casilla 110-V, Valparaíso,
Chile}

\date{\today}


\begin{abstract}
In this paper we discuss the inclusive $J/\psi$ production in proton-proton
collisions from fusion of three pomerons. We demonstrate
that this mechanism gets dominant contribution from the region which
can be theoretically described by CGC/Saturation approach. Numerically,
it gives a substantial contribution to the $J/\psi$ production, and
is able to describe the experimentally observable shapes of the rapidity,
momenta and multiplicity distributions. The latter fact provides a
natural explanation of the experimentally observed enhancement of
multiplicity distribution in $J/\psi$ production.
\end{abstract}

\preprint{TAUP - 3030/18, USM-TH-359}

\maketitle
\tableofcontents{}

\flushbottom

\section{Introduction}

The description of the charmonium hadroproduction remains one of the
long-standing puzzles since its discovery. The large mass $m_{c}$
of the charm quark inspired applications of perturbative methods and
consideration in the formal limit of infinitely heavy quark mass~\cite{Korner:1991kf}.
However, in reality the coupling $\alpha_{s}\left(m_{c}\right)\sim1/3$
is not very small, so potentially some mechanisms suppressed in the
large-$m_{c}$ limit, numerically might give a sizeable contribution.

The Color Singlet Model (CSM) ~\cite{Chang:1979nn,Baier:1981uk,Berger:1980ni}
assumes that the dominant mechanism of charmonia production is the
gluon-gluon fusion supplemented by emission of additional gluon. Early
evaluations in the collinear factorization framework did not agree
with the experimental data at large transverse momenta $p_{T}$ by
several orders of magnitude. The failure of the expansion over $\alpha_{s}$
due to milder suppression of higher order terms at large $p_{T}$~\cite{Brodsky:2009cf,Artoisenet:2008fc}
and co-production of additional quark pairs~\cite{Artoisenet:2007xi,Karpishkov:2017kph}
motivated introduction of the phenomenological Color Octet contributions~\cite{Cho:1995vh,Cho:1995ce}.
The modern NRQCD formulation~\cite{Bodwin:1994jh,Maltoni:1997pt,Brambilla:2008zg,Feng:2015cba,Brambilla:2010cs}
constructs a systematic expansion over the Long Distance Matrix Elements
(LDMEs) of different charmonia states which can be extracted from
fits of experimental data. However, at present extracted matrix elements
depend significantly on the technical details of the fit~\cite{Feng:2015cba,Baranov:2016clx,Baranov:2015laa},
which contradicts expected universality of the extracted LDMEs. At
the same time, it is known that at large $p_{T}$, a sizeable contribution
might come from other mechanisms, like for example gluon fragmentation
into $J/\psi$ or co-production together with other hadrons~\cite{Bodwin:2012xc,Bodwin:2014gia,Braaten:1994xb,Braaten:1995cj}.
The latter findings are partially supported by experimental data on
multiplicity of co-produced charged particles~\cite{PSIMULT,Porteboeuf:2010dw,Sjorstrand:1987,Bartalini:2010,Alice:2012Mult}
which suggest that $J/\psi$ production might get sizeable contribution
from this mechanism.

\begin{figure}[ht]
\centering \leavevmode \includegraphics[width=14cm]{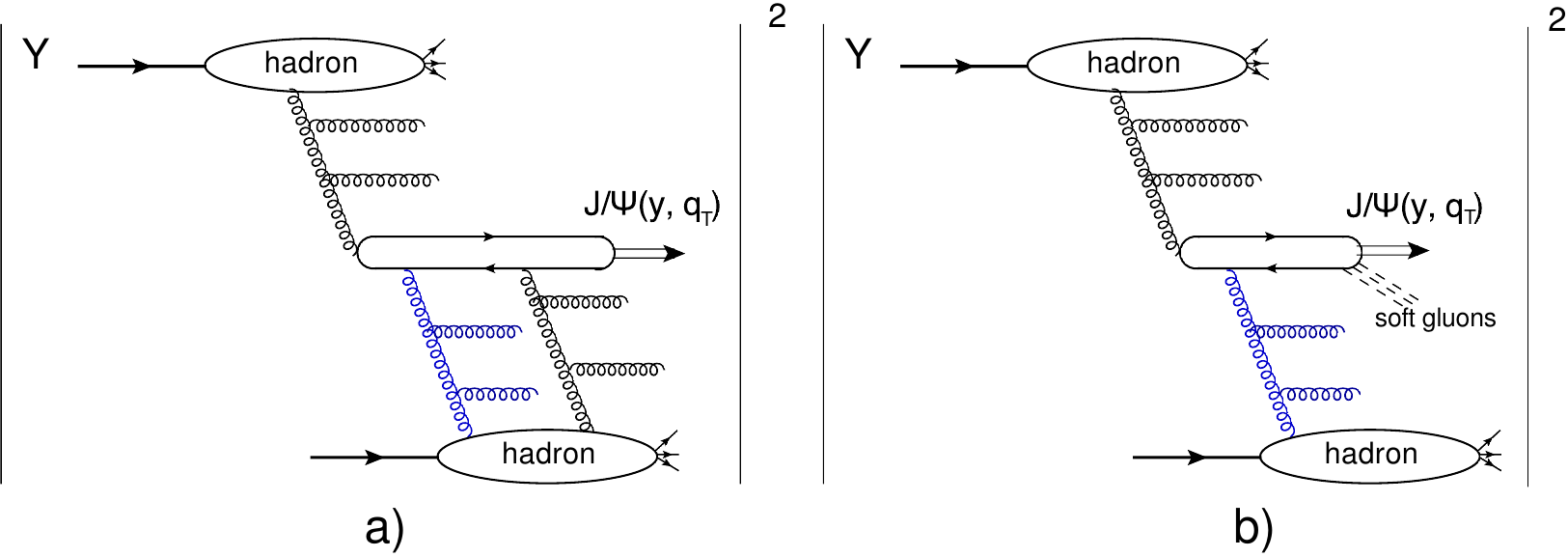}
\caption{Two parton showers contribution to $J/\Psi$ production in hadron-hadron
collisions Fig.~(\ref{2sh}-a) and production of $J/\Psi$ and even
number of soft (non-perturbative) gluons from one parton shower. }
\label{2sh} 
\end{figure}

In this paper we analyze $J/\psi$ in the CGC/Saturation approach,
which incorporates the leading-log contributions from hadron production
in the small-$x$ kinematics, and  takes into account saturation
effects in the region of very small-$x$~\cite{Kovchegov:2012mbw,KOLEB}.
We focus on the mechanism of $J/\psi$ production shown in the diagram
$(a)$ of the Fig.~(\ref{2sh}), when the production of $J/\psi$
occurs from fusion of three gluons, accompanied by the production
of two parton showers~\footnote{We would like to clarify what we mean, saying \textquotedblleft the
production of two parton showers.\textquotedblright{} The parton shower
is the initial state \textquotedblleft cut ladder,\textquotedblright{}
which generates the production of gluons that are almost uniformly
distributed in the entire region of rapidities. As we will see below,
such parton showers correspond to the cut BFKL Pomeron. }. In order to emphasize the role of co-production of other particles
associated with $J/\psi$, below we will follow the terminology used
in some BFKL papers and refer to this mechanism as ``two-parton shower''
contribution. This mechanism differs from the so-called ``single-shower''
mechanism shown in the diagram $(b)$ of the Fig.~(\ref{2sh}). The
latter corresponds to gluon-gluon fusion with emission of additional
(soft) gluon in collinear and $k_{T}$-factorization approaches, and
is a counterpart of CSM mechanism in BFKL framework. We expect that
the two-parton shower mechanism should be dominant for the events
with large multiplicities, exceeding the average multiplicity $\bar{n}$
of the gluon production in the inclusive process. This mechanism is
similar to mechanisms studied earlier in the literature for proton-proton
collisions~\cite{KMRS,Schmidt:2018gep} and for proton-nucleus and
nucleus-nucleus scattering~\cite{KHTU,Schmidt:2018rkw,Kopeliovich:2017jpy}.
For collisions involving heavy nuclei, the two parton shower mechanism
inside heavy nuclei with atomic number $A$ is  dominant due to  factor
$\sim A^{1/3}$, and, because of this, has been comprehensively discussed
during the past decade \cite{KLNT,DKLMT,KMV,GOLEPSI}. On the other
hand, for proton-proton scattering we found only one paper which considers
this process in the $k_{T}$-factorization approach~\cite{MOSA}.

At first sight, the two parton showers mechanism should be suppressed
in comparison with the production of $J/\psi$ in one parton shower
shown in the diagram $(b)$ of the Fig.~(\ref{2sh}-b): in contrast
to $\alpha_{s}$ coupling which appears in the shower-quark vertex,
the emission of soft gluon in Fig.~(\ref{2sh}-b) is not suppressed
by any hard scale and thus does not bring a smallness proportional
to $\bar{\alpha}_{S}$. In the DGLAP approach, it is expected that
such diagrams are of higher twist and are (at least formally) suppressed
by additional powers of hard scale. However, at high energy this suppression
is compensated by enhanced contribution of the second parton shower
(see~\cite{KOLEB} for review). For typical $\langle r\rangle\,\sim\,1\,{\rm GeV}^{-1}$
we get $\bar{\alpha}_{S}\left(4/r^{2}\right)\approx0.2$ so $\bar{\alpha}_{S}G^{{\rm BFKL}}\left(s,\dots\right)\,\propto\bar{\alpha}_{S}\,s^{\Delta_{\small{\rm BFKL}}}\geq1$
at high energies, where $\Delta_{{\rm BFKL}}=4\ln2\bar{\alpha}_{S}$
is the intercept of the BFKL Pomeron\cite{BFKL} and $G^{{\rm BFKL}}\left(s,\dots\right)$
is the Green function of the BFKL Pomeron. The numerical estimates
of Ref.\cite{MOSA} suggests that the considered mechanism yields
about one third of the experimental cross section in the $k_{T}$
factorization approach, though the final estimate suffers significantly
from the uncertainty in digluon PDF modeling, namely the choice of
value of the parameter $\sigma_{{\rm eff}}$(``effective double parton
cross-section''). On the other hand, the diagram $(b)$ in the Fig~(\ref{2sh})
at high energies gets additional suppression due to growth of the
saturation scale $Q_{s}$, which decreases the average dipole size
and suppresses the emission of the extra gluons leading to $ \alpha_S(Q_s)$ suppression.

The main motivation of this paper is to re-visit the estimates of
the contribution of the mechanism of Fig.~(\ref{2sh}-a) in the CGC/Saturation
framework and check if it can reproduce the observed multiplicity
distributions. We demonstrate that: i) this mechanism can be calculated
in CGC/Saturation approach (see Ref.~\cite{Kovchegov:2012mbw} for
a review) since the main contribution comes from the vicinity of the
saturation scale, where we know theoretically the scattering amplitude,
and (ii) it on its own gives a significant contribution to experimentally
observable cross sections. In contrast to Ref.~\cite{MOSA}, we use
a CGC/Saturation framework, and in order to avoid uncertainties related
to digluon distributions, we relate the cross-sections of the $J/\psi$
production to the diffractive production cross-section  known from
DIS.

The paper is organized as follows. In the next section~\ref{sec:2}
we evaluate the contribution of the suggested mechanism in the CGC/Saturation
framework. We re-write this contribution in the coordinate representation
and relate it to the gluon double densities. In Section~\ref{sec:5}
we discuss the interrelation of the suggested process with the diffractive
production of $J/\psi$ in DIS. In the Section~\ref{sec:PhenomenologicalXSection}
we make phenomenological estimates and compare results with experimental
data. The subsection~\ref{sec:6} is dedicated to the estimates of
the total cross section of the process. In this subsection we point
out the differences with Ref.~\cite{MOSA}. In subsections~\ref{sec:7},~\ref{sec:8}
and \ref{sec:9} we consider the momenta, rapidity and multiplicity
distributions of the differential cross-sections. Finally, in the
Section~\ref{sec:10} we draw conclusions. 

\section{Charmonia production in the BFKL  approach}

\label{sec:2} At present, the effective theory of QCD at high energies
exists in two forms: the CGC/saturation approach \cite{GLR,MUQI,MV,MUCD,BK,JIMWLK}
and the BFKL Pomeron calculus~\cite{BFKL,LIREV,GLR,MUPA,BART,BRN,KOLE,LELUR,LMP}.
It has been proven that in general these two approaches are equivalent
in a limited region of the rapidities~\cite{AKLL}. The interpretation
of processes at high energy appears quite different in each approach,
since they have different structural elements.

The CGC/saturation approach, being more microscopic, describes the
high energy interactions in terms of colorless dipoles, their density,
distribution over impact parameters, evolution in energy, etc. The
distinctive feature of this approach is the appearance of saturation
effects which affect the dynamics for parton momenta comparable to
some saturation scale~$Q_{s}$, a new dimensional parameter. The
studies of $J/\psi$ production in this approach may be found in~\cite{KMV}
and by construction include all possible multishower contributions,
although with additional model-dependent assumptions.

The BFKL Pomeron calculus works with BFKL Pomerons and their interactions,
and phenomenologically is similar to the old Reggeon theory \cite{COL}.
This approach is suitable for describing diffractive physics and correlations
in multi-particle production, so we can use the Mueller diagram technique~\cite{MUDI}.
The relation between different processes at high energy are very often
more transparent in this approach, since in addition to the Mueller
diagram technique we can use the AGK cutting rules~\cite{AGK}, which
are useful in spite of the restricted region of their application~\cite{AGKQCD}.
In this paper for the sake of definiteness we use the BFKL Pomeron
calculus as a framework for evaluations.

\begin{figure}[ht]
\centering \leavevmode \includegraphics[width=12cm]{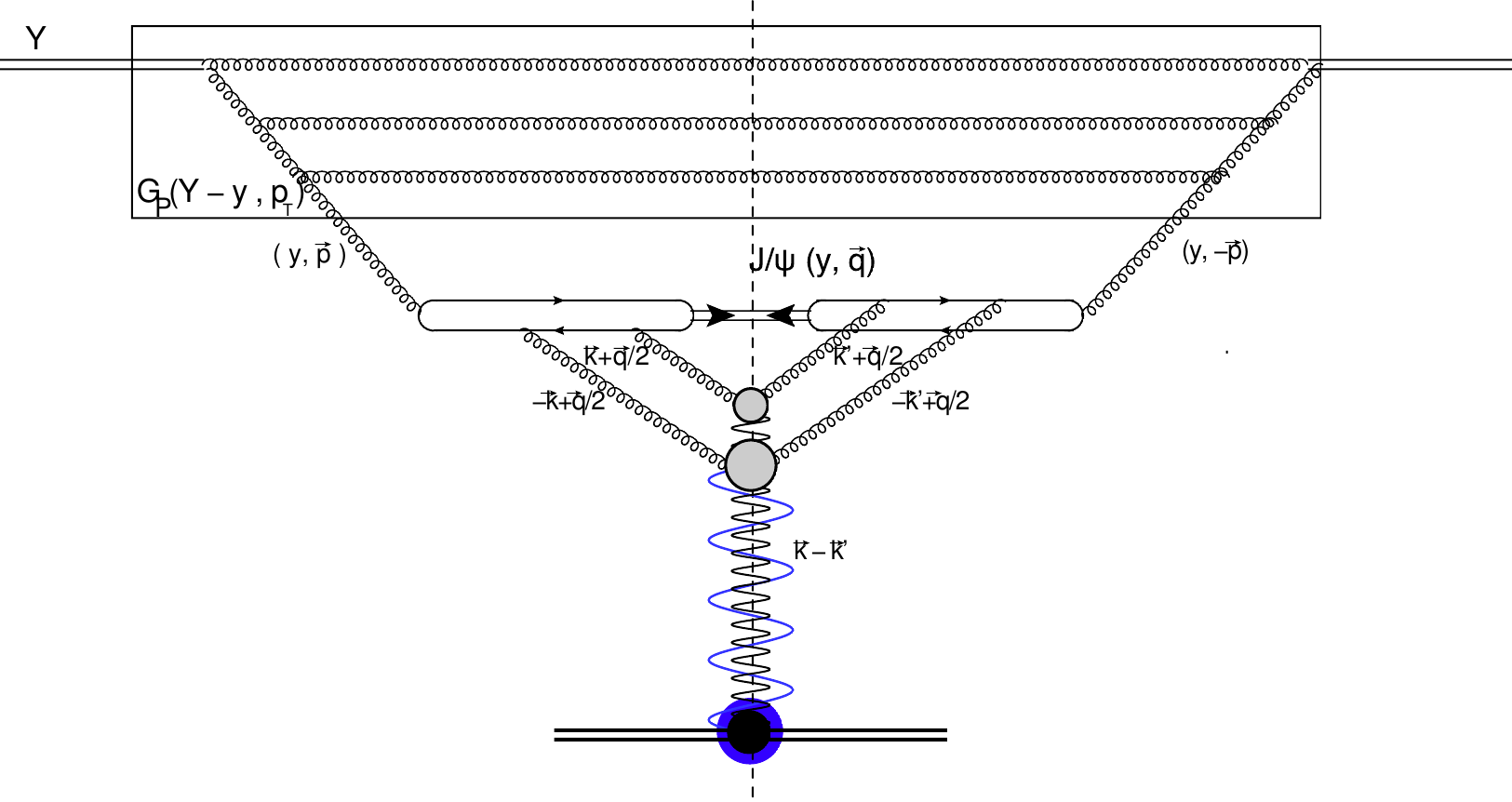}
\caption{The cross-section corresponding to the first diagram of the Fig.~(\ref{2sh})
in the BFKL Pomeron calculus..The vertical wavy lines of different
colors and shape passing through unitarity cut are BFKL Pomerons (described
by the Green functions $G_{I\!\!P}\left(y,\,\boldsymbol{k}_{T}\right)$
in Eq.~(\ref{FD1})), helical lines correspond to the gluons. }
\label{fidi} 
\end{figure}

In the framework of the BFKL Pomeron calculus, the cross-section of
the process shown in the diagram (a)~of the Fig.~(\ref{2sh}) is
described by the exchange of two BFKL Pomerons as shown in Fig.~(\ref{fidi}).
Since we are interested in inelastic $J/\psi$ production, the Pomerons
in Fig.~(\ref{fidi}) are cut Pomerons in which all gluons are produced.
From the unitarity constraints for the elastic amplitude $N^{{\rm BFKL}}$
of the dipole of size $r$, rapidity $Y$ and at the impact parameter
$b$, we have~\cite{Kovchegov:2012mbw} 
\begin{equation}
N_{{\rm cut}}^{{\rm BFKL}}\left(Y,\,r,\,b\right)\,\,\equiv\,\,\,\,\,\,2\,N^{{\rm BFKL}}\left(Y,\,r,\,b\right)\label{UNT}
\end{equation}

Its contribution to the total cross section for $J/\psi$ production\footnote{The color coefficient has been calculated in Ref.\cite{KLNT} and
at large $N_{c}$ it turns out to be equal $C_{F}^{3}$.} is equal to 
\begin{eqnarray}
 &  & \frac{d^{2}\sigma\left(Y,q_{T}\right)}{dy\,d^{2}q_{T}}\,\,=\frac{4\,C_{F}^{3}\bar{\alpha}_{S}^{3}}{(2\pi)^{6}}\int d^{2}k_{T}\,d^{2}p_{T}\,d^{2}Q_{T}\,G_{I\!\!P}^{{\rm cut}}\left(Y-y,\,\boldsymbol{p}_{T},\,0\right)\times\label{FD1}\\
 & \times & I\left(\boldsymbol{k}_{T},\boldsymbol{q}_{T}\right)\,I\left(\boldsymbol{k}'_{T},\boldsymbol{q}_{T}\right)G_{{I\!\!P}}^{{\rm cut}}\left(y,\,\boldsymbol{k}_{T}+\frac{1}{2}\boldsymbol{q}_{T},\boldsymbol{Q}_{T}\right)\,G_{{I\!\!P}}^{{\rm cut}}\left(y,\,-\boldsymbol{k}_{T}+\frac{1}{2}\boldsymbol{q}_{T},\boldsymbol{Q}_{T}\right)\nonumber 
\end{eqnarray}
where $G_{I\!\!P}^{{\rm cut}}$ are the Green functions of the cut
pomerons (it is related to elastic amplitudes $N_{{\rm cut}}^{{\rm BFKL}}$
by Fourier transform), the gluon momenta $\boldsymbol{k}_{T},\,\boldsymbol{p}_{T},\,\boldsymbol{q}_{T}$,
are defined in the Fig.~(\ref{fidi}), $\boldsymbol{Q}_{T}\,=\,\boldsymbol{k}_{T}\,-\,\boldsymbol{k}'_{T}$,
and 
\begin{equation}
I\left(\boldsymbol{k}_{T},\boldsymbol{q}_{T}\right)\,\,=\,\,\int_{0}^{1}dz\int d^{2}r\,\Big(e^{i\frac{1}{2}\boldsymbol{q}_{T}\cdot\boldsymbol{r}}\,\,-\,e^{i\boldsymbol{k}_{T}\cdot\boldsymbol{r}}\Big)\,\,\Psi_{g}\left(p_{T},r;z\right)\,\Psi_{J/\psi}\left(r;\,z\right)\,\,=\,\,F\left(\frac{1}{2}\boldsymbol{q}_{T}\right)\,-\,F\left(\boldsymbol{k}_{T}\right)\label{FD2}
\end{equation}
(the evaluation of the factor $I\left(\boldsymbol{k}_{T},\boldsymbol{q}_{T}\right)$
is discussed in more detail in Appendix~\ref{sec:IFactor}). The
additional factor 4 in Eq.~(\ref{FD9}) which comes from two sources:
2 from the AGK cutting rules~\cite{AGK} and 2 from the fact that
gluon that produces $c\bar{c}$-pair can come from another proton.
In Eq.~(\ref{FD2}) $\Psi_{g}\left(p_{T},\,r\right)$ stands for
the wave function of gluon with virtuality $p_{T}$, transverse quark-antiquark
separation $r$ and the light-cone fraction of the quark $z$, while
$\Psi_{J/\psi}$ is the wave function of $J/\psi$ meson. The amplitude
of the BFKL Pomeron $G_{{I\!\!P}}^{{\rm cut}}\left(y,\,\boldsymbol{k}_{T}\pm\frac{1}{2}\boldsymbol{q}_{T},\,\boldsymbol{Q}_{T}\right)$
can be simplified if we take into account that the $\boldsymbol{Q}_{T}$
dependence of the BFKL Pomeron is determined by the size of the largest
of the interacting dipoles~\footnote{The fact that the $Q_{T}$ dependence is determined by the size of
the largest dipole stem from the general features of the BFKL Pomeron.
Indeed, the eigenfunction of the BFKL Pomeron in the coordinate space
is equal to\cite{LIREV} 
\begin{equation}
N\left(r,r';b\right)\,\,=\,\,\left(\frac{r^{2}\,\,r'^{2}}{\left(\vec{b}-\frac{1}{2}\left(\vec{r}-\vec{r}'\right)\right)^{2}\,\left(\vec{b}-\frac{1}{2}\left(\vec{r}-\vec{r}'\right)\right)^{2}}\right)^{\gamma}\label{BFKLBDEP}
\end{equation}
where $b$ is the conjugate variable to $Q_{T}$. From Eq.~(\ref{BFKLBDEP})
one can see that the typical value of $b$ is of the order of the
largest of $r$ and $r'$. In our process $r'$ is of the order of
$R_{h}$, where $R_{h}$ denotes the radius of the nucleon. The value
of $1/r$ is of the order of the mass of the heavy quark $m_{c}$,
or the saturation scale $Q_{s}$ and, therefore, turns out to be much
larger than $1/R_{h}$, and can be neglected. The dependence on $Q_{T}\approx1/R_{h}$
is described by $S_{h}\left(Q_{T}\right)$ in Eq.~(\ref{FD3}), which
has the non-perturbative origin and. in practice, has to be taken
from the experiment.}, and accounting for Eq.~(\ref{UNT}), may be written as 
\begin{equation}
G_{{I\!\!P}}^{{\rm cut}}\left(y,\,\boldsymbol{k}_{T}\pm\frac{1}{2}\boldsymbol{q}_{T},\,\boldsymbol{Q}_{T}\right)\,\,\approx\,\,2G_{I\!\!P}^{{\rm BFKL}}\left(y,\,\boldsymbol{k}_{T}\pm\frac{1}{2}\boldsymbol{q}_{T},\,\boldsymbol{Q}_{T}=0\right)\,S_{h}\left(Q_{T}\right)\label{FD3}
\end{equation}

The dependence on $Q_{T}$ is described by $S_{h}\left(Q_{T}\right)$
in Eq.~(\ref{FD3}), which has the non-perturbative origin and has
to be taken from the experiment.

Using Eq.~(\ref{FD3}) we can re-write Eq.~(\ref{FD1}) in the form
\begin{eqnarray}
\frac{d\sigma\left(Y,\,Q^{2}\right)}{dy\,d^{2}q_{T}}\,\, & = & \,\,\frac{4\,C_{F}^{3}\,\bar{\alpha}_{S}^{3}}{(2\pi)^{4}}\,\int\frac{d^{2}Q_{T}}{(2\pi)^{2}}\,S_{h}^{2}\left(Q_{T}\right)\,\int d^{2}k_{T}d^{2}p_{T}\,\,G_{I\!\!P}^{{\rm cut}}\left(Y-y,\,p_{T},\,0\right)\,\label{FD5}\\
 & \times & I^{2}\left(\boldsymbol{k}_{T},\boldsymbol{q}_{T}\right)\,G_{{I\!\!P}}^{{\rm BFKL}}\left(y;\boldsymbol{k}_{T}+\frac{1}{2}\boldsymbol{q}_{T},\,0\right)\,G_{{I\!\!P}}^{{\rm BFKL}}\left(y,-\boldsymbol{k}_{T}+\frac{1}{2}\boldsymbol{q}_{T},\,0\right)\nonumber 
\end{eqnarray}

For further evaluations it is very convenient to introduce momenta
$\boldsymbol{p}_{1,2,T}=\pm\boldsymbol{k}_{T}+\frac{1}{2}\boldsymbol{q}_{T}$,
which allow to rewrite Eq.~(\ref{FD5}) as

\begin{eqnarray}
 &  & \frac{d\sigma\left(Y,\,Q^{2}\right)}{dy\,d^{2}q_{T}}\,\,=\,\,\frac{4\,C_{F}^{3}\,\bar{\alpha}_{S}^{2}}{(2\pi)^{4}}\,\int\frac{d^{2}Q_{T}}{(2\pi)^{2}}\,S_{h}^{2}\left(Q_{T}\right)\,x_{g}G\left(x_{g},\,M_{J/\psi}\right)\label{FD8}\\
 &  & \times\int d^{2}p_{1,T}d^{2}p_{1,T}\,\delta^{(2)}\left(\boldsymbol{p}_{1,T}+\boldsymbol{p}_{2,T}-\boldsymbol{q}_{T}\right)\,I^{2}\left(\boldsymbol{p}_{1,T},\boldsymbol{p}_{2,T}\right)\,G_{{I\!\!P}}^{{\rm BFKL}}\left(y;\boldsymbol{p}_{1,T},\,0\right)\,G_{{I\!\!P}}^{{\rm BFKL}}\left(y,\boldsymbol{p}_{2,T},\,0\right)\nonumber 
\end{eqnarray}

where we took the integral over $\boldsymbol{p}_{T}\in\left(0,\,2m_{c}\right)$
using $G_{I\!\!P}^{{\rm cut}}\left(Y-y,\,\boldsymbol{p}_{T},\,0\right)\,\,=\,\,dx\,G\left(x,p_{T}^{2}\right)\Big{/}dp_{T}^{2}$
, $x_{g}G\left(x_{g},\,M_{J/\psi}\right)$ is the gluon structure
function, $x_{g}=M_{J/\psi}e^{y}/\sqrt{s}$, and $Y-y\equiv\ln\left(1/x_{g}\right)$.

Making a Fourier transform, we may rewrite Eq.~(\ref{FD8}) in the
coordinate space as

\begin{eqnarray}
 &  & \frac{d\sigma\left(Y,\,Q^{2}\right)}{dy\,d^{2}q_{T}}\,\,=\,\,\,4\,\int\frac{d^{2}Q_{T}}{(2\pi)^{2}}\,S_{h}^{2}\left(Q_{T}\right)\,\,\,x_{g}G\left(x_{g},\,M_{J/\psi}\right)\,\,\label{FD9}\\
 &  & \times\,\,\,\int_{0}^{1}dz\int_{0}^{1}dz'\int\frac{d^{2}r}{4\pi}\,\frac{d^{2}r'}{4\pi}\,d^{2}b\,\,e^{-i\boldsymbol{q}_{T}\cdot\left(\boldsymbol{b}+\frac{1}{2}(\boldsymbol{r}-\boldsymbol{r}')\right)}\,\,\langle\Psi_{g}\left(r,z\right)\,\Psi_{J/\psi}\left(r,z\right)\rangle\,\langle\Psi_{g}\left(r',z'\right)\,\Psi_{J/\psi}\left(r',z'\right)\rangle\nonumber \\
 &  & \times\,\,\Bigg(N\left(y;\,\boldsymbol{b}\right)\,+\,N\left(y;\,\boldsymbol{b}+\boldsymbol{r}-\boldsymbol{r}'\right)\,-\,N\left(y;\,\boldsymbol{b}+\boldsymbol{r}\right)\,-\,N\left(y;\boldsymbol{b}-\boldsymbol{r}'\right)\Bigg)^{2}\nonumber \\
 &  & \,\,=\,\,4\,\int\frac{d^{2}Q_{T}}{(2\pi)^{2}}\,S_{h}^{2}\left(Q_{T}\right)\,\,\,x_{g}G\left(x_{g},\,M_{J/\psi}\right)\,\,\nonumber \\
 &  & \times\,\,\,\int_{0}^{1}dz\int_{0}^{1}dz'\int\frac{d^{2}r}{4\pi}\,\frac{d^{2}r'}{4\pi}\,d^{2}b\,\,e^{-i\boldsymbol{q}_{T}\cdot\left(\boldsymbol{b}\right)}\,\,\langle\Psi_{g}\left(r,z\right)\,\Psi_{J/\psi}\left(r,z\right)\rangle\,\langle\Psi_{g}\left(r',z'\right)\,\Psi_{J/\psi}\left(r',z'\right)\rangle\,\nonumber \\
 &  & \times\,\,\,\Bigg(N\left(y;b-\frac{1}{2}\left(\boldsymbol{r}-\boldsymbol{r}'\right)\right)\,+\,N\left(y;\boldsymbol{b}+\frac{1}{2}\left(\boldsymbol{r}-\boldsymbol{r}'\right)\right)\,\nonumber \\
 &  & \,\,-\,\,N\left(y;\boldsymbol{b}+\frac{1}{2}\left(\boldsymbol{r}+\boldsymbol{r}'\right)\right)\,\,-\,\,N\left(y;\boldsymbol{b}-\frac{1}{2}\left(\boldsymbol{r}+\boldsymbol{r}'\right)\right)\Bigg)^{2}\nonumber 
\end{eqnarray}
where the amplitudes $N\left(y;\,\boldsymbol{r}_{i}\right)$ are related
to the solutions of the BK equation as $N\left(y;\,\boldsymbol{r}_{i}\right)\,\,=\,\,\int d^{2}b'\,N\left(y;\,\boldsymbol{r}_{i},\,\boldsymbol{b}'\right)$,
and the variable $\boldsymbol{b}$ is a Fourier conjugate of momentum
$\boldsymbol{p}_{1,T}+\text{ }\boldsymbol{p}_{2,T}-\boldsymbol{q}_{T}.$
In what follows, we will also need an expression for the $q_{T}$-integrated
cross-section, which takes a simpler form
\begin{eqnarray}
 &  & \frac{d\sigma\left(Y,Q^{2}\right)}{dy}\,\,=\,\,\,16\,\int\frac{d^{2}Q_{T}}{(2\pi)^{2}}\,S_{h}^{2}\left(Q_{T}\right)\,\,\,x_{g}G\left(x_{g},m_{c}\right)\,\,\label{FD11-1}\\
 &  & \times\,\,\,\int_{0}^{1}dz\int_{0}^{1}dz'\int\frac{d^{2}r}{4\pi}\,\frac{d^{2}r'}{4\pi}\,\,\,\left\langle \Psi_{g}\left(r,z\right)\,\Psi_{J/\psi}\left(r,z\right)\right\rangle \,\left\langle \Psi_{g}\left(r',z'\right)\,\Psi_{J/\psi}\left(r',z'\right)\right\rangle \nonumber \\
 &  & \times\,\,\Bigg(\,N\left(y;\,\frac{\vec{r}+\vec{r}'}{2}\right)\,-\,N\left(y;\,\frac{\vec{r}-\vec{r}'}{2}\right)\Bigg)^{2}\nonumber 
\end{eqnarray}

Finally, we would like to mention that the expressions presented in
this section implicitly assume that each parton cascade is emitted
independently, namely that parton correlations are negligible. In
more general case with nonzero parton correlations, the expression~(\ref{FD5})
should be replaced with 
\begin{eqnarray}
\frac{d\sigma\left(Y,Q^{2}\right)}{dy\,d^{2}q_{T}}\,\, & = & \,\,\frac{4\,C_{F}^{3}\,\bar{\alpha}_{S}^{2}}{(2\pi)^{4}}\,\int\frac{d^{2}Q_{T}}{(2\pi)^{2}}\,\int d^{2}k_{T}d^{2}p_{T}\,\,G_{I\!\!P}^{{\rm cut}}\left(Y-y,\,p_{T},\,0\right)\,\nonumber \\
 & \times & I^{2}\left(\boldsymbol{k}_{T},\boldsymbol{q}_{T}\right)\,\rho^{(2)}\left(y;\boldsymbol{k}_{T}+\frac{1}{2}\boldsymbol{q}_{T};\,y,-\boldsymbol{k}_{T}+\frac{1}{2}\boldsymbol{q}_{T};\,\boldsymbol{Q}_{T}\right)
\end{eqnarray}
where we replaced the product of pomeron propagators with the double
transverse momentum densities $\rho^{(2)}$ defined as

\begin{eqnarray}
 &  & \rho^{(2)}\left(x_{1},\boldsymbol{p}_{1,T};x_{2},\boldsymbol{p}_{2,T},\boldsymbol{Q}_{T}\right)\,=\,\,\label{eq:rho2}\\
 &  & \,\left\langle P\left|\Bigg\{ a^{+}\left(x_{1},\boldsymbol{p}_{1,T}+\frac{1}{2}\boldsymbol{Q}_{T};b\right)\,a^{+}\left(x_{2},\boldsymbol{p}_{2,T}-\frac{1}{2}\boldsymbol{Q}_{T};c\right)\,a\left(x_{2},\boldsymbol{p}_{2,T}+\frac{1}{2}\boldsymbol{Q}_{T};c\right)\,a\left(x_{1},\boldsymbol{p}_{1,T}-\frac{1}{2}\boldsymbol{Q}_{T};b\right)\Bigg\}\right|P\right\rangle \nonumber 
\end{eqnarray}
where $\left(x_{1},p_{1,T}\right)$ and $\left(x_{2},p_{2,T}\right)$
are the light-cone and transverse momenta of the partons in the cascade,
\footnote{The definition~(\ref{eq:rho2}) is closely related to digluon PDFs
introduced in~\cite{Diehl:2011yj}, though we have to mention that
the probabilistic interpretation strictly speaking can be discussed
only for $Q_{T}=0$. As we will see below, in final expressions for
the $J/\psi$ production we will need only this case. } $|P\rangle$ is the Fock state of colliding hadrons, $a^{+}$ and
$a$ denote the creation and annihilation operators for gluons that
have longitudinal momentum $x_{i}$ and transverse momentum $p_{i,T}$,
$c_{i}$ are the color indexes. However, at present there is no experimental
evidence that such correlations are large, for this reason in what
follows we will use a simpler expressions~(\ref{FD5}). Similarly,
in coordinate space the Eq.~(\ref{FD9}) can be extended as 
\begin{eqnarray}
 &  & \frac{d\sigma\left(Y,Q^{2}\right)}{dy\,d^{2}q_{T}}\,\,=\,\,4\,\int\frac{d^{2}Q_{T}}{(2\pi)^{2}}\,\,x_{g}G\left(x_{g},2\,m_{c}\right)\,\,\label{FD10}\\
 &  & \times\,\,\,\int_{0}^{1}dz\int_{0}^{1}dz'\int\frac{d^{2}r}{4\pi}\,\frac{d^{2}r'}{4\pi}\,d^{2}b\,\,e^{-i\boldsymbol{q}_{T}\cdot\boldsymbol{b}},\langle\Psi_{g}\left(r,z\right)\,\Psi_{J/\psi}\left(r,z\right)\rangle\,\langle\Psi_{g}\left(r',z\right)\,\Psi_{J/\psi}\left(r',z\right)\rangle\nonumber \\
 &  & \times\,\,\Bigg(\rho^{(2)}\left(y;\boldsymbol{b}-\boldsymbol{r}_{d};y;\boldsymbol{b}-\boldsymbol{r}_{d},Q_{T}\right)\,+\,\rho^{(2)}\left(y;\boldsymbol{b}+\boldsymbol{r}_{d},y;\boldsymbol{b}+\boldsymbol{r}_{d},Q_{T}\right)\,+\,\rho^{(2)}\left(y;\boldsymbol{b}+\boldsymbol{r}_{s},y;\boldsymbol{b}+\boldsymbol{r}_{s},Q_{T}\right)\nonumber \\
 &  & \,+\,\rho^{(2)}\left(y;\boldsymbol{b}-\boldsymbol{r}_{s},y;\boldsymbol{b}-\boldsymbol{r}_{s},Q_{T}\right)\,+\,2\,\rho^{(2)}\left(y;\boldsymbol{b}-\boldsymbol{r}_{d},y;\boldsymbol{b}+\boldsymbol{r}_{s},Q_{T}\right)\,-\,2\,\rho^{(2)}\left(y;\boldsymbol{b}+\boldsymbol{r}_{d},y;\boldsymbol{b}+\boldsymbol{r}_{s},Q_{T}\right)\,\nonumber \\
 &  & -\,\rho^{(2)}\left(y;\boldsymbol{b}-\boldsymbol{r}_{d},y;\boldsymbol{b}-\boldsymbol{r}_{s},Q_{T}\right)\,-\,2\,\rho^{(2)}\left(y;\boldsymbol{b}+\boldsymbol{r}_{d},y;\boldsymbol{b}+\boldsymbol{r}_{s},Q_{T}\right)\,-\,2\,\rho^{(2)}\left(y;\boldsymbol{b}+\boldsymbol{r}_{d},y;\boldsymbol{b}-\boldsymbol{r}_{s},Q_{T}\right)\,\nonumber \\
 &  & +\,2\,\rho^{(2)}\left(y;\boldsymbol{b}+\boldsymbol{r}_{s},y;\boldsymbol{b}-\boldsymbol{r}_{s},Q_{T}\right)\Bigg)\nonumber 
\end{eqnarray}
where $\rho^{(2)}\left(x,\,\boldsymbol{r};\,x,\,\boldsymbol{r}';\,\boldsymbol{Q}_{T}\right)$
is the double parton density in the coordinate representation, $\boldsymbol{r}_{s}\,=\,\frac{1}{2}(\boldsymbol{r}+\boldsymbol{r}')$
and $\boldsymbol{r}_{d}\,=\,\frac{1}{2}(\boldsymbol{r}-\boldsymbol{r}')$.

While for numerical estimates we could use parameterizations of the
amplitude~$N$ available from the literature, we would like to minimize
dependence on parameterization and make a few model-independent estimates.
One of the important parameters for understanding the  small-$x$
dynamics is the saturation scale~$Q_{s}$ and its product on characteristic
size of the dipoles $\langle r\rangle$ in the process. The saturation
has a mild dependence on energy~\cite{DKLN} and in the kinematics
of interest for our studies ($\sqrt{s}\in\left(1.9,\,7\right)$ TeV)
its values are $Q_{s}^{2}=0.7-0.9\,{\rm GeV}^{2}$. The typical size
of the dipole essential in our process might be estimated as~\footnote{See Appendix~\ref{sec:IFactor} and~\cite{PSIWF} for paramertizations
of the wave functions}
\begin{equation}
\left\langle r^{2}\right\rangle \,=\,\frac{\int_{0}^{1}dz\int\,\frac{d^{2}r}{4\pi}\,\,\,r^{2}\left\langle \Psi_{g}\left(r,z\right)\,\Psi_{J/\psi}\left(r,z\right)\right\rangle }{\int_{0}^{1}dz\int\frac{d^{2}r}{4\pi}\,\,\,\left\langle \Psi_{g}\left(r,z\right)\,\Psi_{J/\psi}\left(r,z\right)\right\rangle }\approx0.76\,{\rm GeV}^{-2}\label{TYPR}
\end{equation}
so the product $\tau\,\,\equiv\,\,\left\langle r^{2}\right\rangle \,Q_{s}^{2}\approx0.5\,...\,0.7$.
Contrary to the large-$m_{c}$ expectations, this number is not very
small and corresponds to the dynamics in the vicinity of the saturation
scale \cite{MUT,GOLEPSI}. The scattering amplitude is well known
in this region (see (\ref{eq:NParam}) below), so this implies minimal
model-dependence in our estimates. 

\section{Interrelation with the diffractive $J/\psi$ production in DIS}

\label{sec:5}
\begin{figure}[ht]
\centering \leavevmode \includegraphics[width=12cm]{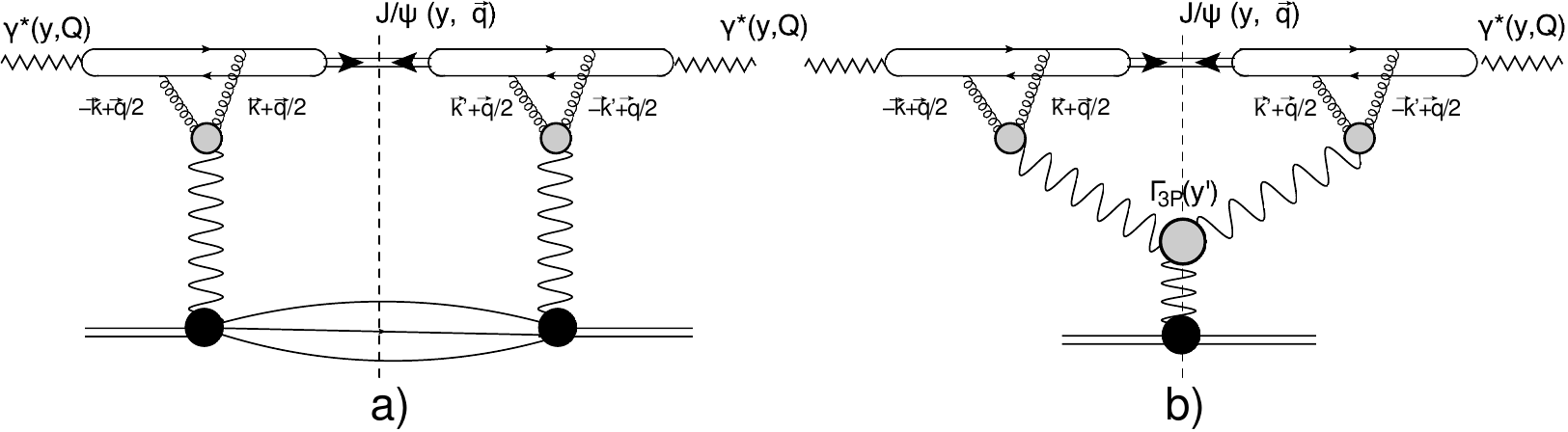}
\caption{\label{fig:Two-photon}Two main contributions to the diffractive production
of $J/\Psi$ meson. The elastic contribution (diagram $(a))$ contributes
mainly to production of hadrons with small total mass, while the inelastic
contribution (diagram $(b)$) is the source of hadrons with large
total mass. }
\label{dispsi} 
\end{figure}

The cross-section of the $J/\psi$ production in the
small-$x$ kinematics is closely related to the diffractive $J/\Psi$
production in DIS, and this relationship is useful to fix the unknown
non-perturbative factor $\sim\int d^{2}Q_{T}\,S_{h}^{2}(Q_{T})$ which
appears in~(\ref{FD8},\ref{FD9}). In the BFKL picture, there are
elastic and inelastic contributions to diffractive $J/\psi$ production,
shown in the left and right panels of the Figure~\ref{fig:Two-photon}.
Taking into account approximate equality of the elastic and inelastic
contributions~\cite{H1PSI,H1ELPSI}, we may focus on evaluation of
the diagram $(a)$ of the Fig.~(\ref{dispsi}). For fixing the prefactor
$\sim\int d^{2}Q_{T}\,S_{h}^{2}(Q_{T})$, it is sufficient to consider
only the $q_{T}$-integrated cross-section $d\sigma_{{\rm diff}}/dy$,
which is given by~(see Fig.~(\ref{dispsi})): 

\begin{eqnarray}
 &  & \frac{d\sigma_{{\rm diff}}\left(y,\,Q^{2},\,\sqrt{s_{\gamma^{*}p}}\right)}{dy}\,\,=\,\,\,\int\frac{d^{2}Q_{T}}{(2\pi)^{2}}\,S_{h}^{2}\left(Q_{T}\right)\int_{0}^{1}dz\int_{0}^{1}dz'\int\frac{d^{2}r}{4\pi}\,\frac{d^{2}r'}{4\pi}\times\label{FD11}\\
 & \times & \left\langle \Psi_{\gamma^{*}}\left(r,z\right)\,\Psi_{J/\psi}\left(r,z\right)\right\rangle \,\left\langle \Psi_{\gamma^{*}}\left(r',z'\right)\,\Psi_{J/\psi}\left(r',z'\right)\right\rangle N\left(y,\,r^{2}\right)\,N\left(y,\,r'^{2}\right).\nonumber 
\end{eqnarray}
In the vicinity of the saturation scale, which gives the dominant
contribution, the CGC/saturation approach predicts that the amplitude
$N$ has a form

\begin{equation}
N\left(y,\,r^{2},\,0\right)\approx{\rm Const}_{s}\left(r^{2}\,Q_{s}^{2}(x)\right)^{\bar{\gamma}},\label{eq:NParam}
\end{equation}
 where $\bar{\gamma}\approx0.63$~\cite{Kovchegov:2012mbw}, $x\approx M_{J/\psi}e^{-y}/\sqrt{s}$,
and $Q_{s}^{2}(x)$ is the saturation scale. The Eq.~(\ref{FD11})
in this case takes the form 
\begin{eqnarray}
 &  & \frac{d\sigma_{{\rm diff}}\left(y,\,Q^{2},\,\sqrt{s_{\gamma^{*}p}}\right)}{dy}\,\,=\mbox{Const}_{{\rm s}}^{2}\,\,\,\int\frac{d^{2}Q_{T}}{(2\pi)^{2}}\,S_{h}^{2}\left(Q_{T}\right)\int_{0}^{1}dz\int_{0}^{1}dz'\int\frac{d^{2}r}{4\pi}\,\frac{d^{2}r'}{4\pi}\,\label{FD110}\\
 & \times & \,\,\left\langle \Psi_{\gamma^{*}}\left(r,\,z\right)\,\Psi_{J/\psi}\left(r,\,z\right)\right\rangle \,\left\langle \Psi_{\gamma^{*}}\left(r',\,z'\right)\,\Psi_{J/\psi}\left(r',\,z'\right)\right\rangle \left(Q_{s}^{2}\left(x\right)\,r^{2}\right)^{\bar{\gamma}}\,\,\left(Q_{s}^{2}\left(x\right)\,r'^{2}\right)^{\bar{\gamma}},\nonumber 
\end{eqnarray}
which allows to factorize all the energy- and rapidity dependence,
\begin{equation}
\frac{d\sigma_{{\rm diff}}\left(y,\,Q^{2},\,\sqrt{s_{\gamma^{*}p}}\right)}{dy}\sim\,\left[Q_{s}^{2}\left(x\right)\right]^{2\bar{\gamma}}
\end{equation}
and in this way facilitate scaling from HERA to Tevatron and LHC energies.
Similarly, for the gluon induced elastic $J/\psi$ production in the
vicinity of the saturation scale (see Ref.\cite{GOLEPSI}) we may simplify~(\ref{FD11-1})
to
\begin{eqnarray}
 &  & \frac{d\sigma\left(y,\,Q^{2},\,\sqrt{s_{\gamma^{*}p}}\right)}{dy}\,=\,16\,\,\,\mbox{Const}_{{\rm s}}^{2}\int\frac{d^{2}Q_{T}}{(2\pi)^{2}}\,S_{h}^{2}\left(Q_{T}\right)\,\,\,x_{g}G\left(x_{g},m_{J/\Psi}\right)\,\,\label{FD12}\\
 &  & \times\,\,\,C_{F}^{2}\,\,\int_{0}^{1}dz\int_{0}^{1}dz'\int\frac{d^{2}r}{4\pi}\,\frac{d^{2}r'}{4\pi}\,\,\,\left\langle \Psi_{g}\left(r,z\right)\,\Psi_{J/\psi}\left(r,z\right)\right\rangle \,\left\langle \Psi_{g}\left(r',z'\right)\,\Psi_{J/\psi}\left(r',z'\right)\right\rangle \nonumber \\
 &  & \,\Bigg\{\left(\int d^{2}b\,Q_{s}^{2}\left(x,\,b\right)\right)^{2\bar{\gamma}}\left(\left(\frac{\vec{r}+\vec{r}'}{2}\right)^{2\,\bar{\gamma}}\,-\,\left(\frac{\vec{r}-\vec{r}'}{2}\right)^{2\,\bar{\gamma}}\right)^{2}\Bigg\}\nonumber 
\end{eqnarray}
which also has a factorizable dependence on energy and rapidity, 
\begin{equation}
\frac{d\sigma\left(y,\,Q^{2}\right)}{dy}\sim x_{g}\,G\left(x_{g}\,M_{J/\psi}\right)\,\left[Q_{s}^{2}\left(x\right)\right]^{2\bar{\gamma}}.
\end{equation}

Taking into account that the wave functions of the proton and gluon
are proportional to each other in the leading order of pQCD~\cite{DKLMT},
\begin{equation}
\Psi_{\gamma^{*}}\left(r,\,z\right)\,\,=\,\,\frac{\sqrt{2N_{c}}}{g}\frac{2}{3}eN_{c}\Psi_{g}\left(r,\,z\right)
\end{equation}
we may expect the proportionality of the diffractive and inclusive
cross-sections 
\begin{equation}
\frac{d\sigma\left(y,\,Q^{2}\right)/dy}{d\sigma_{{\rm diff}}\left(y,\,Q^{2}\right)/dy}\Bigg{|}_{Q=m_{J/\psi},y\to Y_{0}}\,\,\approx\frac{9}{2}\,x_{g}G\left(x_{g},\,M_{J/\psi}\right)\,\mathcal{R}\,C_{F}^{2}\,\frac{\alpha_{S}}{\alpha_{{\rm e.m,.}}N_{c}^{3}}\label{R}
\end{equation}
where 
\begin{align}
\mathcal{R} & =I_{1}/I_{2}\approx1.27,\\
I_{1} & =4\int_{0}^{1}dz\int_{0}^{1}dz'\int\frac{d^{2}r}{4\pi}\int\frac{d^{2}r'}{4\pi}\left\langle \Psi_{\gamma^{*}}\left(r,z\right)\,\Psi_{J/\psi}\left(r,z\right)\right\rangle \,\left\langle \Psi_{\gamma^{*}}\left(r',z'\right)\,\Psi_{J/\psi}\left(r',z'\right)\right\rangle \\
 & \times\left[\left(\frac{\vec{r}+\vec{r}'}{2}\right)^{2\bar{\gamma}}-\left(\frac{\vec{r}-\vec{r}'}{2}\right)^{2\bar{\gamma}}\right]^{2},\nonumber \\
I_{2} & =\int_{0}^{1}dz\int_{0}^{1}dz'\int\frac{d^{2}r}{4\pi}\int\frac{d^{2}r'}{4\pi}\left\langle \Psi_{\gamma^{*}}\left(r,z\right)\,\Psi_{J/\psi}\left(r,z\right)\right\rangle \,\left\langle \Psi_{\gamma^{*}}\left(r',z'\right)\,\Psi_{J/\psi}\left(r',z'\right)\right\rangle \left(r^{2}r'^{2}\right)^{\bar{\gamma}}
\end{align}
and for estimates of parameter $\mathcal{R}$   we used the parameterizations
of the wave functions given in Appendix~\ref{sec:IFactor}. The data
on diffractive production are available from H1 and ZEUS experiments~\cite{H1PSI,ZEUSPSI,H1ELPSI},
and they show that in HERA kinematics the elastic and inelastic contributions
(diagrams $a$ and $b$ in the Fig.~\ref{dispsi}, respectively) are
approximately equal. As a consequence, for $W\equiv\sqrt{s}\approx30\,{\rm GeV}$
($x_{g}\approx0.01$) the diffractive cross-section
\[
\frac{d\sigma_{{\rm diff}}}{dy}\Bigg{|}_{Q=m_{J/\psi},y\to Y_{0}}\sim20\,{\rm nb}.
\]

This allows to to use Eq.~(\ref{R}) for estimate of the hadroproduction
cross-section at $\sqrt{s}\approx30$~GeV,
\begin{equation}
\left.\frac{d\sigma}{dy}\right|_{x_{g}\approx10^{-2}}\approx0.83\,{\rm \mu b}.\label{GXS}
\end{equation}

In the next section we will extrapolate it with the help of the small-$x$
evolution up to LHC energies. 

\section{Phenomenological estimates}

\label{sec:PhenomenologicalXSection}

\subsection{Total $J/\Psi$ production cross section }

\label{sec:6}

As we demonstrated in Section~\ref{sec:2}, the typical size $\langle r\rangle$~of
$Q\bar{Q}$ dipole is small when the saturation scale $Q_{s}(x)\lesssim m_{c}$\textbf{~}(see
Eq.~(\ref{TYPR})\textbf{)}, and for this reason from ~(\ref{FD12})
we expect that the cross-section at central rapidities should scale
with energy as 
\begin{equation}
\left.\frac{d\sigma}{dy}\right|_{y=0}\sim x_{g}G\left(x_{g},\,M_{J/\psi}\right)\left(Q_{s}^{2}\left(x\right)\right)^{2\bar{\gamma}}.
\end{equation}
For numerical estimates we assume that the saturation scale $Q_{s}^{2}$
scales as~\cite{MUT}
\begin{equation}
Q_{s}^{2}\left(x\right)=Q_{s}^{2}\left(x_{0}\right)\,\left(\frac{x_{0}}{x}\right)^{\lambda}.\label{QS1-2}
\end{equation}

with $\lambda\approx0.2-0.3$~\cite{Watt:2007nr,Kowalski:2003hm,RESH}
and $x\approx M_{J/\psi}e^{-y}/\sqrt{s}$.
 In case of Tevatron and
LHC kinematics, this leads to the cross-section estimates given in
the Table~ \ref{tab:Comparison}. In these estimates we use that the evolution does not change the ratio between  the elastic and inelastic contributions which correspond to two terms in the solution to the evolution equation for $\rho^{(2)}$ (see Appendix B and  Fig.~(\ref{2cont})).

 As we can see, the suggested mechanism
gives a significant contribution to the total cross-section, though
in view of inherent uncertainties of the CGC/Saturation approach we
cannot make more accurate estimate about the fraction of charmonia
produced via this mechanism. 
\begin{table}
\begin{tabular}{|c|c|c|}
\hline 
 & Theoretical estimates  & Experiment\tabularnewline
\hline 
$\sqrt{s}\approx1.96$ TeV  & 2.1-2.6~$\mu$b  & 2.38~ $\mu$b~\cite{PSICDF}~\tabularnewline
\hline 
$\sqrt{s}\approx7$ TeV  & 3.8-5.6~$\mu$b  & 5.8~$\mu$b~\cite{Abelev:2012gx}\tabularnewline
\hline 
\end{tabular}

\caption{\label{tab:Comparison}Comparison of the theoretical estimates with
experimental data for the cross-section $d\sigma/dy$ at central rapidities.
Theoretical estimates correspond to values of parameter $\lambda\in(0.2,\,0.3)$
(lower and upper values respectively). In the last column we've quoted
the data on \emph{prompt} $J/\psi$ production. The cross-section
$d\sigma/dy$ at Tevatron was extracted dividing the total cross-section
$\sigma_{{\rm tot}}(|y|<0.6)$ by the width of the bin $\Delta y\sim1.2$. }
\end{table}

As we have mentioned earlier, our estimates exceed the result of Ref.~\cite{MOSA},
where similar contribution was found approximately twice smaller.
Below we would like to analyze the reasons which might be responsible
for this discrepancy. It was assumed in~\cite{MOSA} that the digluon
distribution is proportional to a direct product of independent gluon
distributions, with a phenomenological multiplicative pre-factor usually
described by the so-called effective cross-section $\sigma_{{\rm eff}}$.
In the CGC/Saturation picture, each gluon effectively is replaced
by a pomeron, so the factorized model would correspond to the diagram
$(a)$ in the Fig.~(\ref{dispsi}). As we discussed earlier, the
inelastic contributions (diagram $(b)$ in the same Figure) yields
a numerically comparable contribution which would correspond to corrections
breaking the factorized form. This additional inelastic contribution
might be one of the reasons of the strong channel dependence of the
phenomenological effective cross-section $\sigma_{{\rm eff}}$~\footnote{This phenomenological parameter manifests significant dependence on
channel used to extract it and varies between 5 and 25 ${\rm mb}$
(see Ref.\cite{DPI} for a review).}. In our approach, we fix the the only unknown constant $\int d^{2}Q_{T}\,S^{2}\left(Q_{T}\right)$
from experimental data at HERA and evolve the cross-section as prescribed
by small-$x$ evolution, thus taking into account both diagrams of
the Fig.~(\ref{dispsi}). The main source of uncertainty in this
procedure is the energy dependence of the saturation scale $Q_{s}^{2}(x)$,
namely the choice of coefficient $\lambda$ in~(\ref{QS1-2}). Theoretical
estimates of this parameter significantly depend on the choice of
the scale in the running coupling constant $\bar{\alpha}_{s}$, and
lead to estimates $\lambda\approx0.3-0.4$~\cite{CLMP1}, whereas
phenomenological estimates of this parameter restrict it to the range
$\lambda\in(0.2,\,0.3)$~\cite{Kovchegov:2012mbw}. We can see that
it results in up to fifty per cent uncertainty in the theoretical
estimates shown in the Table~\ref{tab:Comparison}.

\subsection{Transverse momentum distribution}

\label{sec:7} Qualitatively the main features of $q_{T}$-distribution
may be understood from (\ref{FD9}). For small momenta $\frac{1}{2}q_{T}\lesssim Q_{s}\left(x\right)$,
we can safely use the behaviour of the Pomeron Green function in the
vicinity of the saturation momentum~\cite{MUT} \textbf{
\begin{eqnarray}
G_{{I\!\!P}}^{{\rm BFKL}}\left(y;\,\boldsymbol{k}_{T}\,\pm\,\frac{1}{2}\boldsymbol{q}_{T},\,\boldsymbol{Q}_{T}=0\right) & \approx & {\rm const}\left(\frac{Q_{s}^{2}\left(x\right)}{\left(\boldsymbol{k}_{T}\,\pm\,\frac{1}{2}\boldsymbol{q}_{T}\right)^{2}}\right)^{\bar{\gamma}},\qquad{\rm small}\,\boldsymbol{q}_{T}.
\end{eqnarray}
}

In the kinematic region $\frac{1}{2}q_{T}\gtrsim k_{T}$ the scattering
amplitude becomes sensitive to dynamics at shorter distances, which
is described in perturbative QCD. In this region we may use a parameterization~(\ref{eq:NParam}),
which is a BFKL prediction for small dipoles with $r\lesssim Q_{s}^{-1}\left(x\right)$~\cite{Kovchegov:2012mbw}. 

The two approaches are valid for different values of $q_{T}$ and
thus complement each other. The choice of the threshold scale $q_{0}$
between them  is somewhat arbitrary, yet we expect that  $q_{0}$
should be of order $\sim(1-2)\,m_{J/\psi}$. 
 
\begin{figure}[ht]
\centering \leavevmode \includegraphics[width=12cm]{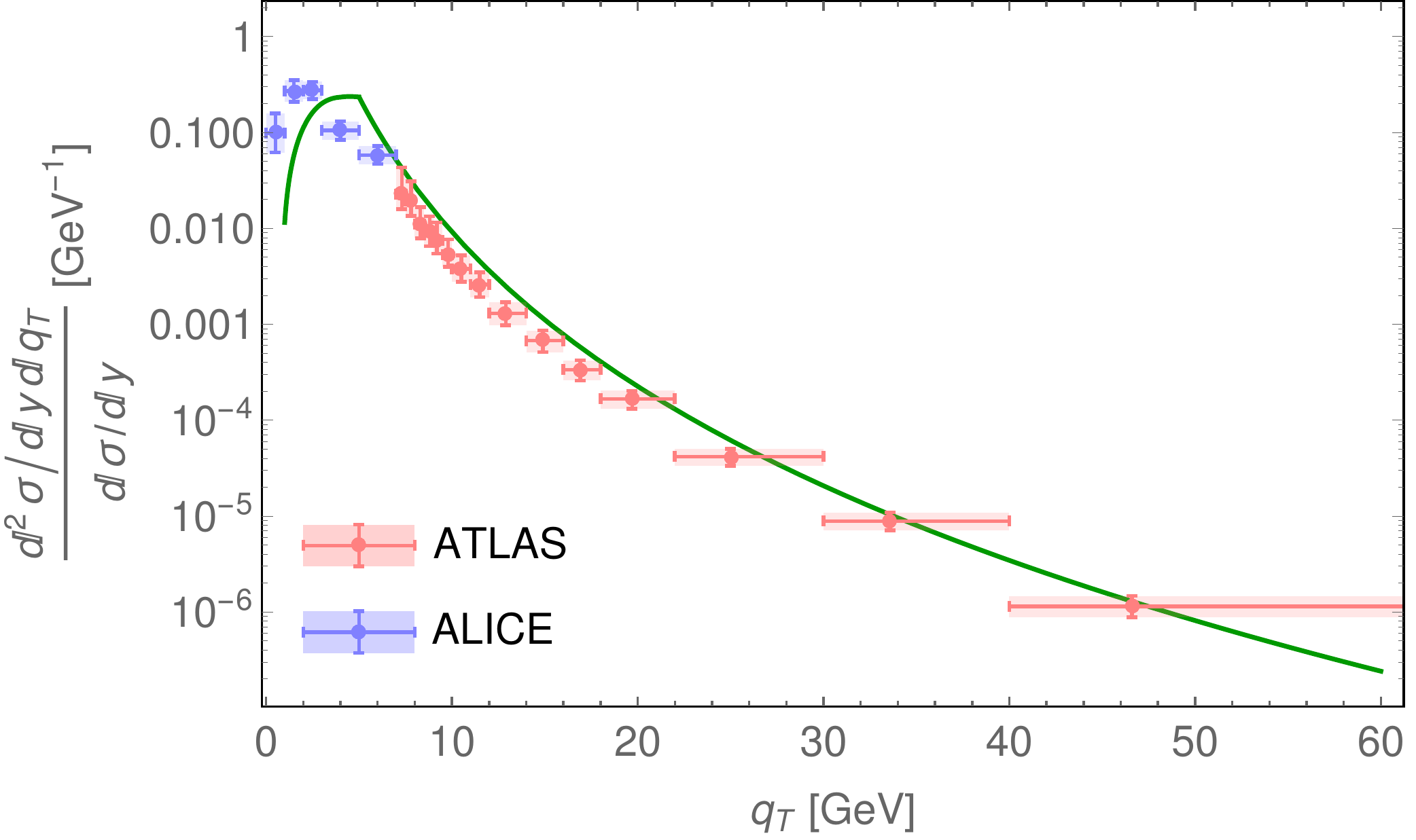} \caption{The shape of the $q_{T}$ distribution of produced $J/\psi$ at central
rapidities. The threshold scale $q_{0}$ is chosen as $q_{0}\sim$5\,~GeV.
The experimental data are taken from Refs.~\cite{PSIATLAS,PSIALICE}.}
\label{pt} 
\end{figure}

In Fig.~(\ref{pt}) we compare model predictions for the shape of
the $q_{T}$-dependence with experimental data from~\cite{PSIATLAS,PSIALICE}.
In order to avoid the above-mentioned global uncertainty in normalization,
we consider the normalized ratio $\left(d^{2}\sigma/dy\,dq_{T}\right)/\left(d\sigma/dy\right)$.
The cusp on the curve near the threshold scale $q_{T}=q_{0}$ can
be smoothed out by more relaxed conditions.

\subsection{Rapidity distribution}

\label{sec:8} 
\begin{figure}[ht]
\centering \leavevmode \includegraphics[width=12cm]{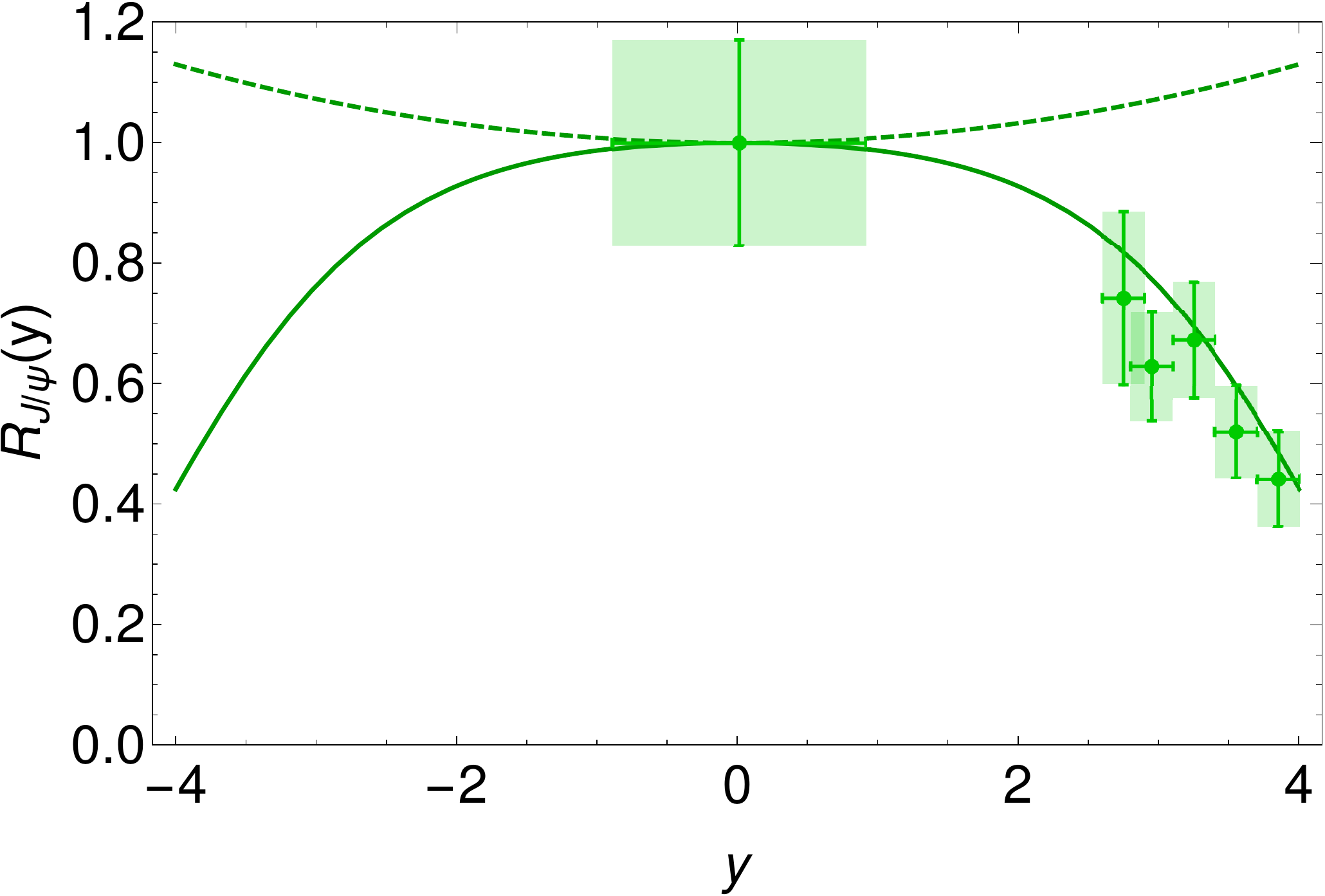}
\caption{Rapidity distribution of produced $J/\psi$. The ratio $R_{J/\psi}(y)$
is defined as $R_{J/\psi}=\left(d\sigma_{J/\psi}/dy\right)/\left(d\sigma_{J/\psi}/dy\right)_{y=0}$.
The dashed curve corresponds to Eq.~(\ref{YDIST1}) with BFKL-style
parameterization~\ref{eq:YDIST3A} for gluon densities, the solid
line takes into account $\sim(1-x)^{5}$-endpoint factors as described
in the text. The data are taken from Ref.\cite{PSIALICE}.}
\label{disty} 
\end{figure}

For numerical estimates in the previous sections we
used the leading log approximation (LLA) for the BFKL Pomeron Green
function, assuming additionally that mass $m_{c}$ is large, so that
we could use a small-$r$ approximation. The shape of the rapidity
distribution in this approximation has a very simple form, which follows
directly from Eq.~(\ref{FD12}): 
\begin{align}
\frac{d\sigma_{J/\psi}\left(y\right)/dy}{d\sigma_{J/\psi}\left(y\right)/dy|_{y=0}}\,\, & =\underbrace{\Bigg(\frac{Q_{s}^{2}\left(y^{*}-y\right)\left(Q_{s}^{2}\left(y^{*}+y\right)\right)^{2}}{Q_{s}^{2}\left(y^{*}\right)\left(Q_{s}^{2}\left(y^{*}\right)\right)^{2}}\Bigg)^{\bar{\gamma}}}_{{\rm Fig.~(\ref{mult})-a}}\,\,+\,\,\underbrace{\Bigg(\frac{Q_{s}^{2}\left(y^{*}+y\right)\left(Q_{s}^{2}\left(y^{*}-y\right)\right)^{2}}{Q_{s}^{2}\left(y^{*}\right)\left(Q_{s}^{2}\left(y^{*}\right)\right)^{2}}\Bigg)^{\bar{\gamma}}}_{{\rm Fig.~(\ref{mult})-b}}\,\,=\,\,\cosh\left(\lambda\bar{\gamma}y\right),\label{YDIST1}\\
 & y^{*}\,\,=\,\,-\,\ln\left(\sqrt{\frac{m_{{J/\psi}}^{2}+q_{T}^{2}}{s}}\right),
\end{align}
since all dependence on rapidity is concentrated in the energy dependence
of the saturation scale (\ref{QS1-2}) which contributes to the cross-section~(\ref{FD12})
multiplicatively, and the gluon density in prefactor. The latter in
the small-$x$ kinematics is expected to have a simple power-like
behaviour 
\begin{equation}
x_{g}G\left(x_{g},\,m_{J/\psi}^{2}\right)\,\propto\,\left(Q_{s}^{2}\left(x_{g}\right)/m_{J/\psi}^{2}\right)^{\bar{\gamma}}.\label{eq:YDIST3A}
\end{equation}
However, such simple parameterization is valid only at central rapidities
(near $y=0$), and as could be seen from the Figure~\ref{disty},
outside this region ($|y|\gtrsim1$) mismatches the experimental cross-section
even on the qualitative level. This happens because at very forward
or very backward kinematics we should add in~(\ref{YDIST1}) additional
factor 
\[
\sim\left(1-x\right)^{5},\qquad x\,=\,e^{-y^{*}\pm y}
\]
per each gluon/pomeron in order to have correct endpoint behaviour
of the gluon densities in the $x\to1$ limit~\cite{BBSH}. As we
can see from the Figure~\ref{disty}, the model gives a reasonable
description of the rapidity dependence, which might be interpreted
as confirmation of the applicability of the small-$r$ approximation.

\subsection{Multiplicity dependence}

\label{sec:9}
\begin{figure}[ht]
\centering \leavevmode \includegraphics[width=12cm]{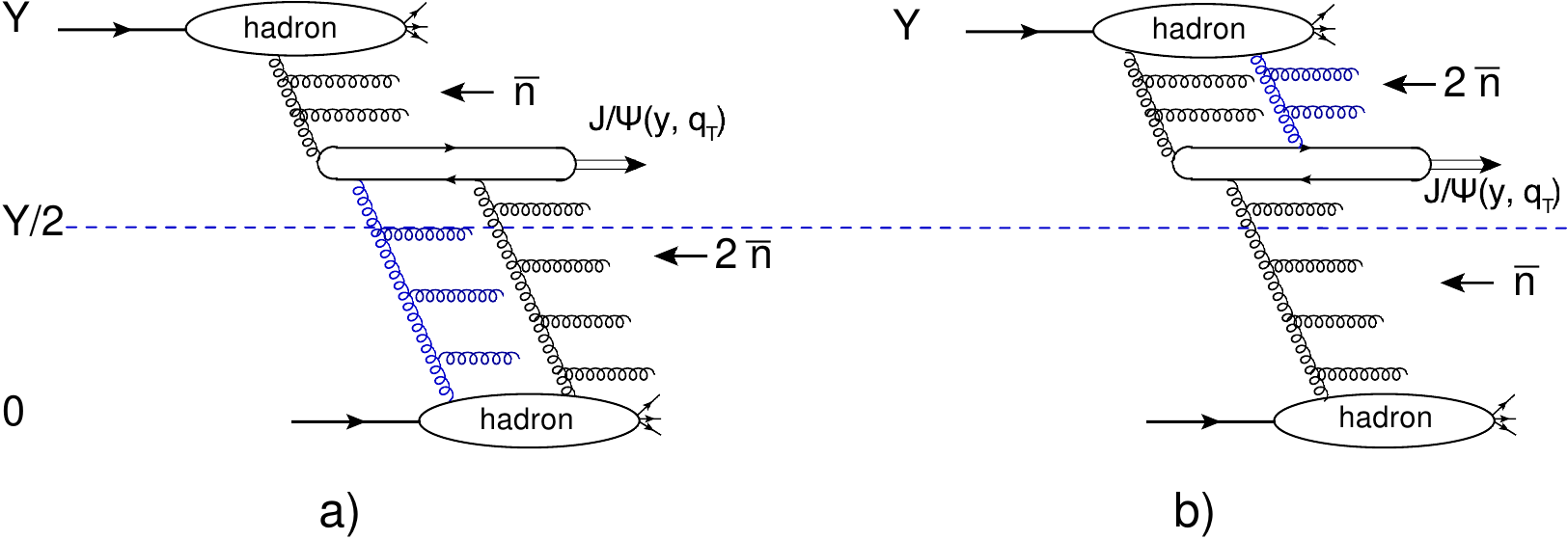}
\caption{Multiplicities in the $J/\psi$ production, accompanied by two parton
showers.}
\label{mult} 
\end{figure}

The $J/\psi$ production accompanied by two parton showers
occurs in the events with larger multiplicity than the average multiplicity
$\bar{n}$ in the inclusive hadron production~~\cite{PSIMULT,Alice:2012Mult}.
The mechanism suggested in this paper provides a natural explanation
of this enhancement. The cross-section includes contributions of additional
parton shower as shown in Fig.~(\ref{mult})-a and Fig.~(\ref{mult})-b,
which enhances the observed multiplicity. Technically, the dependence
on multiplicity $n$ of the produced particles affects the cross-section
through the increase of the number of the particles in a parton shower
and change of the value of the saturation scale, which is proportional
to the density of produced gluons (see.\cite{KOLEB,KLN} for more
details). Assuming that for hadron-hadron collisions the area of interaction
does not depend on multiplicity, the saturation scale is linearly
proportional to the number of particles in the shower~\cite{LERE},
\begin{equation}
Q_{s}^{2}\left(y,\,n\right)\,\,=Q^{2}\left(y,\,\bar{n}\right)\frac{n}{\bar{n}},\label{QSN}
\end{equation}
where $\bar{n}$ is the average multiplicity at $y^{*}=0$ ($\bar{n}\approx$6.5
at $W=13\,{\rm TeV}$~\cite{ALICEINCL}). As we discussed earlier,
the product of dipole size on saturation scale $\left\langle r^{2}\right\rangle Q^{2}\left(y,\,\bar{n}\right)$
is close to one, for this reason the multiplicity events with $n/\bar{n}\gg1$
probe a deep saturation regime, where approximation~(\ref{FD12})
is not valid, and we should use~(\ref{FD11-1}) with phenomenological
parameterization of the dipole amplitude $N$.

\begin{figure}[ht]
\centering \includegraphics[width=9cm]{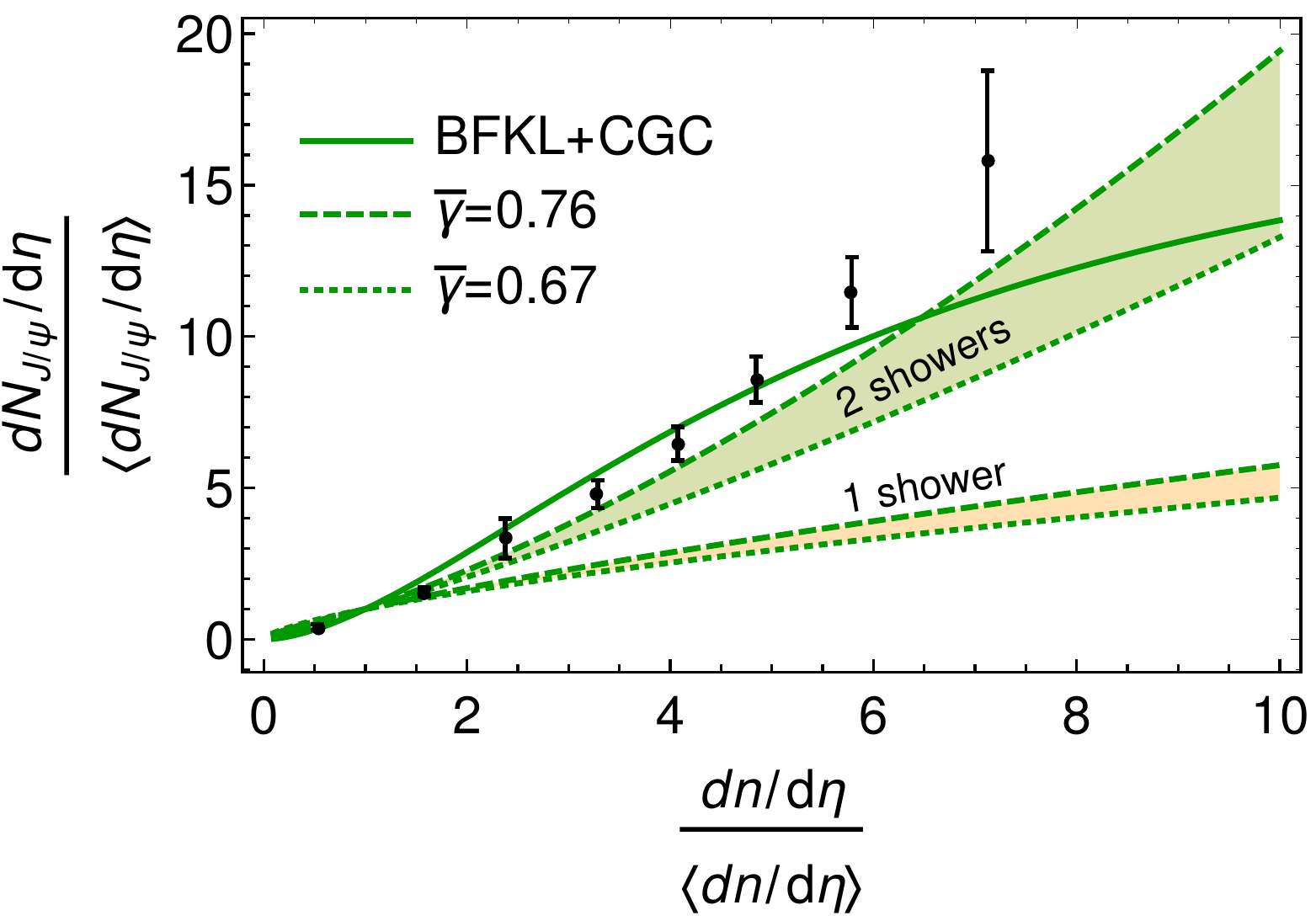}\includegraphics[width=9cm]{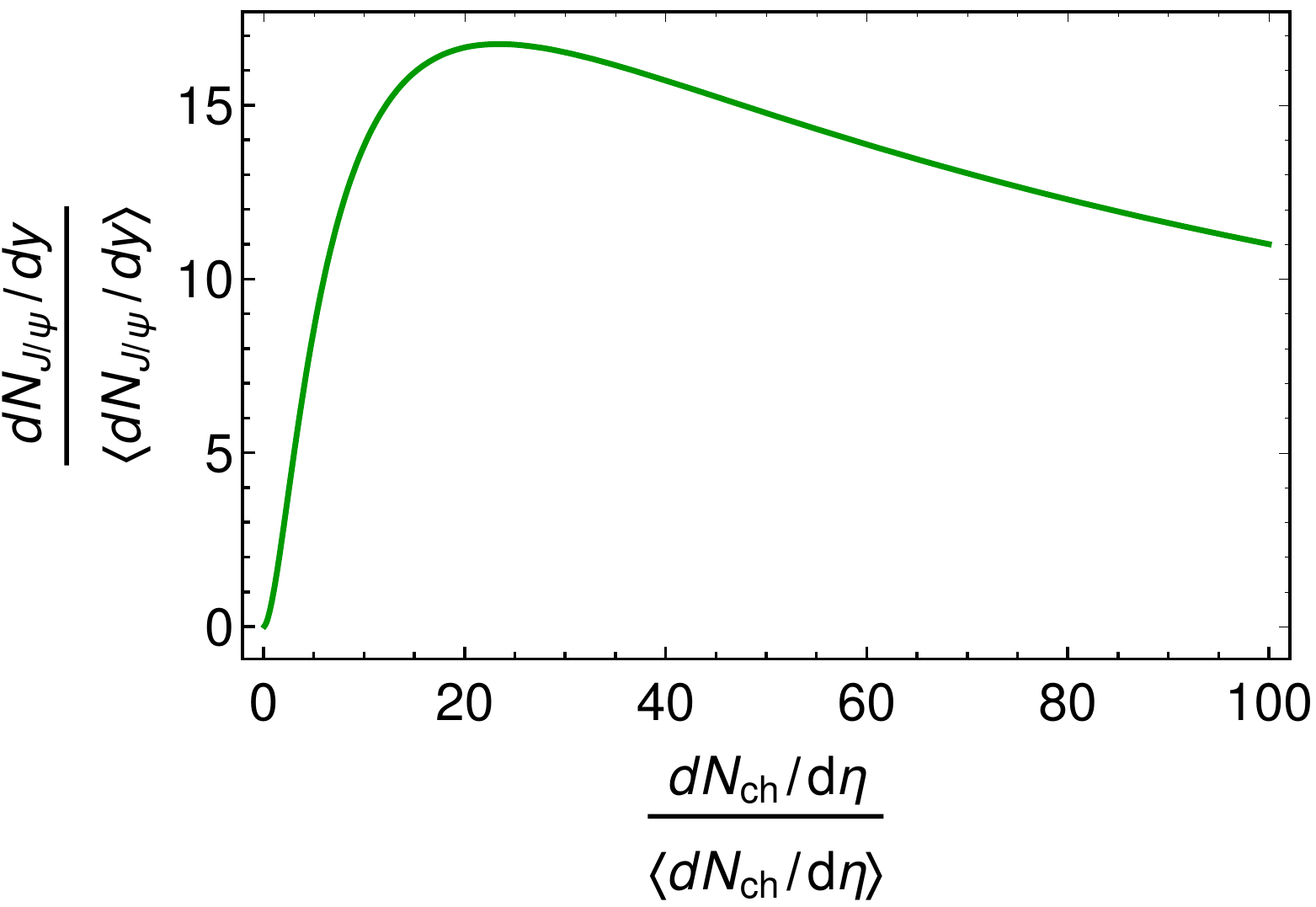}
\caption{Left plot: comparison of the multiplicity distribution with the experiment~\cite{PSIMULT}.
Solid line: result of evaluations with~(\ref{FD11-1}) using CGC
parameterization for the dipole amplitude and saturation scale adjusted
according to~(\ref{QSN}). The upper band marked ``2-showers''
stands for the estimates with approximate expression~(\ref{XSN})
and values of $\bar{\gamma}$ varied in the range $\bar{\gamma}\in\left(0.67,\,0.76\right)$,
as implemented in phenomenological parameterizations. Similarly, the
lower color band marked with label ``1-shower'' stands for multiplicity
distribution evaluated in single-shower mechanism shown in the Fig.~(\ref{2sh})-b.
Right plot: large-multiplicity extension of the solid curve from the
left plot (result of evaluations with~(\ref{FD11-1}) using CGC parameterization).}
\label{commult} 
\end{figure}

In the Figure~\ref{commult} we plot the self-normalized multiplicity
distribution evaluated with (\ref{FD11-1}), using CGC parameterization
for the dipole amplitude and saturation scale adjusted according to~(\ref{QSN}).
We can see that agreement with experimental data from ALICE~\cite{PSIMULT}
is reasonable. At sufficiently small multiplicities, $dN_{J/\psi}/dy$
is increasing as a function of $dN_{{\rm ch}}/d\eta$, however, as
can be seen from the right plot, at larger multiplicities in deep
saturation regime it starts decreasing. This behaviour might be understood
from the structure of the last line in~(\ref{FD11-1}): the dipole
amplitude $N$ saturates (approaches asymptotically a constant)~\cite{LETU},
for this reason the difference $\left[N\left(y;\,\frac{\vec{r}+\vec{r}'}{2},\,0\right)\,-\,N\left(y;\,\frac{\vec{r}-\vec{r}'}{2},\,0\right)\right]^{2}$
gets suppressed. For the sake of completeness in the left panel of
the Figure~\ref{commult} we also plotted the ratio evaluated with
the simplified parameterization~(\ref{eq:NParam}). Taking into account
that contributions of left and right diagrams in the Figure~(\ref{mult})
contribute with relative weights $\sim\left(Q_{s}^{2}\left(y,n\right)\right)^{2\bar{\gamma}}$
and \,$\sim\left(Q_{s}^{2}\left(y,n\right)\right)^{\bar{\gamma}}$
respectively, we may obtain for the self-normalized multiplicity dependence
of contribution in Fig.~(\ref{mult}) a simple expression 
\begin{equation}
\frac{\frac{d\sigma_{J/\Psi}}{dy}\Big{|}_{\small{\rm fixed\,n}}\,\,\,\,\,}{\langle\frac{d\sigma_{J/\Psi}}{dy}\rangle\Big{|}_{\small{\rm sum\,over\,n}}}\,\,=\,\,\frac{1}{1+\kappa}\Bigg(\kappa\left(\frac{n}{\bar{n}}\right)^{\bar{\gamma}}+\,\,\left(\frac{n}{\bar{n}}\right)^{2\,\bar{\gamma}}\Bigg)\label{XSN}
\end{equation}
where the coefficient $\kappa=\left(Q_{s}^{2}\left(Y-y\right)/Q_{s}^{2}\left(y\right)\right)^{\bar{\gamma}}\approx\,\,e^{\bar{\gamma}\,\lambda\left(Y-2y\right)}$
reflects the relative suppression of the contribution of the right
diagram in the Fig.~(\ref{mult}) with respect to the left diagram
for different rapidities (at central rapidities used for comparison
with data $\kappa\approx1$). While the LO CGC/Saturation predicts
a value $\bar{\gamma}\approx0.63$~, the phenomenological fits~\cite{Watt:2007nr,RESH}
favor slightly higher values of $\bar{\gamma}$, for this reason we
varied this parameter in the range $\bar{\gamma}\in\left(0.67,\,0.76\right).$
We also plotted the estimates of single-shower mechanism shown in
the right part of the Figure~(\ref{2sh}). Within model uncertainty,
we can see that the experimental data support the main hypothesis
of the paper that the $J/\psi$ production accompanied by the production
of two parton showers gives a significant contribution.

\section{Conclusions}

\label{sec:10} In this paper we analyzed in the CGC/Saturation framework
the $J/\psi$ hadroproduction, accompanied by production of two parton
showers. We demonstrated that this mechanism gives a large contribution
to the total cross section of $J/\psi$ production, as well as to
the transverse momenta and multiplicity distribution of produced $J/\psi$.
The experimental data~\cite{PSIMULT} on multiplicity distributions
of produced $J/\psi$ seem to favor this hypothesis (compared to conventional
mechanisms estimated in the CGC/Saturation approach). This mechanism
has no suppression of the order of $\bar{\alpha}_{S}$ at high energies
compared to one parton shower contribution. We show that it alone
can describe the shape of the $p_{T}$- and rapidity dependence of
the cross section. As we commented in detail in section 5, our results
are approximately twice larger than similar evaluation of Ref.\cite{MOSA}
and for this reason give a significant contribution to the experimental
data. However, inherent uncertainties of the CGC/Saturation approach
preclude more precise estimates of its fraction. Since the main contribution
to the cross section of this process stems from the vicinity of the
saturation scale, the behaviour of the scattering amplitude in this
region is largely under theoretical control, and thus has minor dependence
on phenomenological parameterizations of the dipole amplitude.

We believe that our studies will bring attention to the contribution
of multiparton densities  in production of $J/\psi$.

\section{Acknowledgements}

We thank our colleagues at Tel Aviv university and UTFSM for encouraging
discussions. This research was supported by the BSF grant 2012124,
by Proyecto Basal FB 0821(Chile), Fondecyt (Chile) grant 1180118 and
by CONICYT PIA grants ACT1406 and ACT1413.

\appendix

\section{Evaluation of the factor~$I\left(\boldsymbol{p}_{1,T},\boldsymbol{p}_{2,T}\right)$}

\label{sec:IFactor} 
\begin{figure}[ht]
\centering \leavevmode \includegraphics[width=8cm]{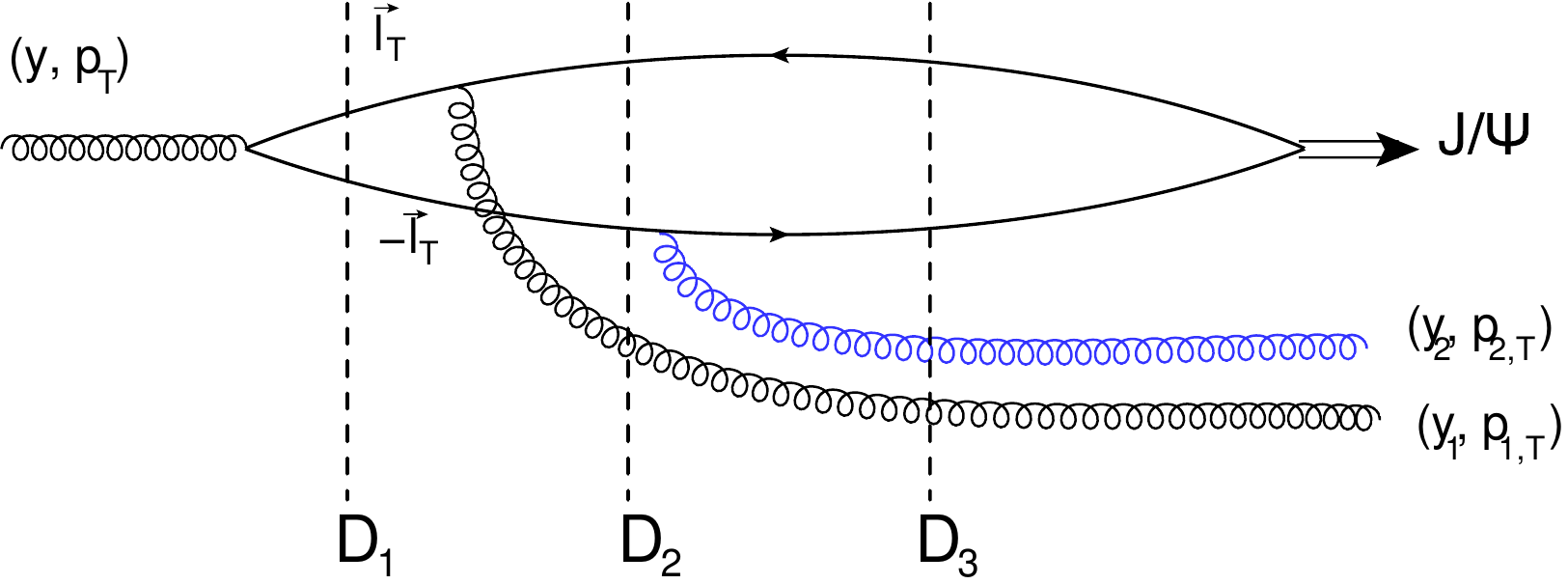}
\caption{The diagram for $I\left(\boldsymbol{p}_{1,T},\boldsymbol{p}_{2,T}\right)$.}
\label{2gdi} 
\end{figure}

In this section we would like to comment briefly the evaluation of
the function $I\left(\boldsymbol{k}_{T},\boldsymbol{q}_{T}\right)$
in~(\ref{FD2}) in the light cone perturbative QCD~\cite{KOLEB}. 

We choose the polarization vectors of gluons $\epsilon_{\mu}^{\lambda}(p)$
in the light-cone gauge as 
\begin{equation}
\epsilon_{\mu}^{\alpha}(p)\,\,=\,\,\Bigg(0,\,\frac{2\,\boldsymbol{\epsilon}_{\perp}^{\alpha}\cdot\boldsymbol{p}_{T}}{\eta\cdot p},\,\boldsymbol{\epsilon}_{\perp}^{\alpha}\Bigg);~~~~~\eta\,=\,(0,1,0,0)~~~~~~\mbox{and}~~~~\boldsymbol{\epsilon}_{\perp}^{\pm}\,\,=\,\,\frac{1}{\sqrt{2}}(\pm1,i)\label{POLVEC}
\end{equation}

so according to the general rules of light cone perturbation theory~(LCPT)
~\cite{MUDI,KOLEB} the wave function of the $\bar{Q}Q$ pair after
interaction with a pair of gluons with momenta ($y_{1,2},\,\boldsymbol{p}_{1,2;\,T}$)
is given by 

\begin{eqnarray}
 &  & \Psi^{(1,1)}\left(\boldsymbol{q}_{T},z;\boldsymbol{p}_{1,T},y_{1};\boldsymbol{p}_{2,T},y_{2}\right)\,=\label{ID4}\\
 &  & \,g^{2}\lambda^{a}\lambda^{b}\frac{\boldsymbol{p}_{1,T}\cdot\boldsymbol{\epsilon}_{1}^{\alpha}}{p_{1,T}^{2}}\frac{\boldsymbol{p}_{2,T}\cdot\boldsymbol{\epsilon}_{2}^{\alpha}}{p_{2,T}^{2}}\Bigg(\Psi_{g}\left(\boldsymbol{l}_{T},z\right)-\Psi_{g}\left(\boldsymbol{l}_{T}+\boldsymbol{p}_{1,T},z\right)-\Psi_{g}\left(\boldsymbol{l}_{T}+\boldsymbol{p}_{2,T},z\right)\,\,+\Psi_{g}\left(\boldsymbol{l}_{T}+\boldsymbol{p}_{1,T}+\boldsymbol{p}_{2,T},z\right)\Bigg)\nonumber 
\end{eqnarray}
where $\lambda^{a}$ is the Gell-Mann matrix, $\boldsymbol{\epsilon}^{\alpha}$
is the polarization vector of the gluon with the helicity $\alpha$,
and $\Psi_{g}\left(\boldsymbol{l}_{T},\,z\right)$ is the wave function
of $\bar{Q}Q$ pair formed after gluon splitting into quark-antiquark
pair with transverse momentum $\boldsymbol{l}_{T}$ and light-cone
fraction $z$. 

In the coordinate space the convolution of~(\ref{ID4}) with the
wave function of $J/\psi$ yields
\begin{equation}
I\left(\boldsymbol{p}_{1,T},\boldsymbol{p}_{2,T}\right)\,\,=\,\,\int_{0}^{1}dz\,\int d^{2}r\left\langle \Psi_{g}\left(r,\,z\right)\,\Psi_{J/\Psi}\left(r,z\right)\right\rangle \,\,e^{-\frac{1}{2}\boldsymbol{q}_{T}\cdot\boldsymbol{r}}\Bigg(1\,\,-\,\,e^{i\boldsymbol{p}_{1,T}\cdot\boldsymbol{r}}\Bigg)\,\Bigg(1\,\,-\,\,e^{i\boldsymbol{p}_{2,T}\cdot\boldsymbol{r}}\Bigg)\label{I}
\end{equation}
which agrees with~(\ref{FD2}) if we change notations of momenta
to $\boldsymbol{p}_{1,2\,T}=\frac{\boldsymbol{q}_{T}}{2}\pm\boldsymbol{k}_{T}$.
In view of the heavy mass of the charm quark, in what follows we will
use leading order perturbative results for the gluon wave functions
$\Psi_{g}$ , as well as phenomenological parameterization of the $J/\psi$
wave function available from the literature~\cite{PSIWF}, so the
convolution of the two objects explicitly takes the form 
\begin{eqnarray}
\left\langle \Psi_{g}\left(r,z\right)\,\Psi_{J/\Psi}\left(r,z\right)\right\rangle _{T} & =\, & \frac{g}{\pi\sqrt{2\,N_{c}}}\frac{1}{z(1-z)}\left(m_{c}^{2}\,K_{0}\left(\epsilon\,r\right)\,\phi_{T}\left(r,z\right)\,\,-\,\,\left(z^{2}+\left(1-z\right)^{2}\right)\,\epsilon\,K_{1}\left(\epsilon\,r\right)\partial_{r}\phi_{T}\left(r,z\right)\,\right);\label{PSIT}\\
\left\langle \Psi_{g}\left(r,z\right)\,\Psi_{J/\Psi}\left(r,z\right)\right\rangle _{L} & =\, & \frac{g}{\pi\sqrt{2\,N_{c}}}2p_{T}z(1-z)\,K_{0}\left(\epsilon\,r\right)\left(m_{J/\Psi}\phi_{L}(r,z)\,\,+\,\,\frac{m_{c}^{2}-\nabla_{r}^{2}}{m_{J/\Psi}\,z\,(1-z)}\phi_{L}(r,z)\right)\,;\label{PSIL}\\
\phi_{T,L}(r,z) & = & \,\,N_{T,L}\,z(1-z)\exp\Bigg(-\frac{m_{c}^{2}\,{\cal R}^{2}}{8\,z\,(1-z)}\,-\,\frac{2\,z\,(1-z)\,r^{2}}{{\cal R}^{2}}\,\,+\,\,\frac{m_{c}^{2}\,{\cal R}^{2}}{2}\Bigg);\label{PHILT}\\
\epsilon^{2} & = & \,p_{T}^{2}\,z\,(1-z)\,\,+\,\,m_{c}^{2}\label{EPSILON}
\end{eqnarray}

\section{Evolution equation for the double gluon density}

\begin{figure}[ht]
\centering \leavevmode \includegraphics[width=8cm]{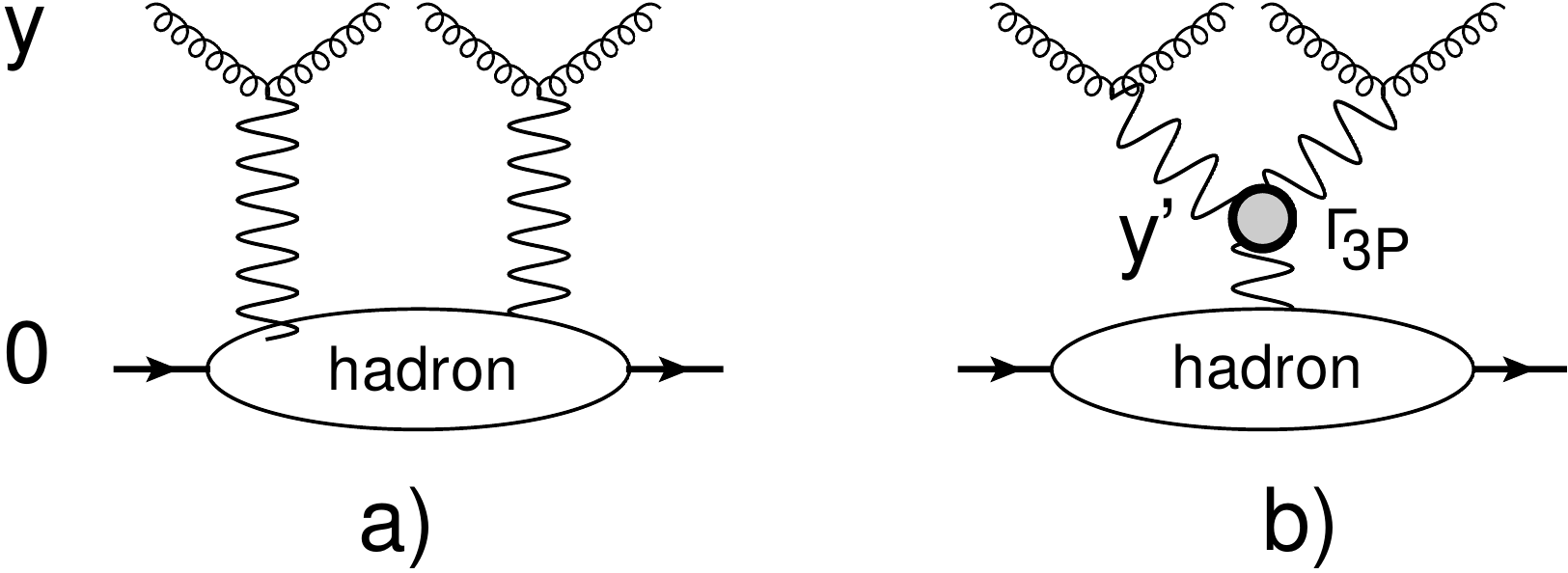}
\caption{Graphical illustration of the solution ~(\ref{SOL}). The diagram
$(a)$ with the exchange of two BFKL Pomerons (denoted by wavy lines)
corresponds to the first term containing $\tilde{\rho}_{{\rm in}}^{(2)}\left(\gamma_{1},\gamma_{2}\right)$;
the diagram $(b)$ which includes the triple Pomeron vertex $\Gamma_{3{I\!\!P}}$
corresponds to the second term containing $\tilde{\rho}_{{\rm in}}^{(1)}\left(\gamma_{1}+\gamma_{2}\right)$.
Helicoidal lines denote gluons. }
\label{2cont} 
\end{figure}

\textbf{} In Ref.\cite{LELU} it is proven that the evolution equations
for all partonic densities $\rho^{(n)}\left(\{\vec{r}_{i},\vec{b}_{i}\}\right)$
are the linear BFKL evolution equations. The non-linear corrections
are essential for the scattering amplitude (see, for example, the
Balitsky-Kovchegov equation for dipole scattering amplitude\cite{BK}
with nuclei) but they do not give a contribution to the evolution
equation for multi-gluon densities. Referring our reader to Ref.\cite{LELU}
for the proof and details, we present here the resulting evolution
equation for $r_{1}^{2}\,r_{2}^{2}\,\int d^{2}b\,\rho_{A}^{(2)}\left(x,\vec{r};x,\vec{r}';\vec{b}_{T}\right)\,=\,\tilde{\rho}^{(2)}\left(Y-Y_{0},\vec{r}_{1},\vec{r}_{2}\right)$:

\begin{eqnarray}
 &  & \frac{\partial\,\tilde{\rho}^{(2)}(Y-Y_{0};\vec{r}_{1},\vec{r}_{2})}{\bar{\alpha}_{S}\,\partial\,Y}\,=\label{EVEQ}\\
 &  & \,\sum_{i=1}^{2}\int\frac{d^{2}r'}{2\pi}\frac{1}{\left(\vec{r}_{i}-\vec{r}'\right)^{2}}\Bigg\{2\tilde{\rho}^{(2)}(Y-Y_{0};\vec{r}',\vec{r}_{i+1})\,-\,\frac{r_{1}^{2}}{r'^{2}}\tilde{\rho}^{(2)}(Y-Y_{0};\vec{r}_{1},\vec{r}_{2})\Bigg\}\,\,+\,\,\tilde{\rho}^{(1)}\left(Y-Y_{0},\vec{r}_{1}+\vec{r}_{2}\right)\nonumber 
\end{eqnarray}
where $\left(\vec{r}_{1}+\vec{r}_{2}\right)^{2}\int d^{2}b\,\rho^{(1)}\left(Y-Y_{0},\vec{r}_{1}+\vec{r}_{2},\,\vec{b}\right)\,\,=\,\,\tilde{\rho}^{(1)}\left(Y-Y_{0},\vec{r}_{1}+\vec{r}_{2}\right)$.
The two terms in Eq.~(\ref{EVEQ}) have clear physical meaning: the
first one is the evolution of two parton showers, while the second
describes the production of two gluon in one parton shower (see Fig.~(\ref{2cont})).

The solution takes the following form (see Ref.\cite{GOLEPSI} and
references therein):

\begin{eqnarray}
 &  & \rho^{(2)}\left(Y,r_{1},Y,r_{2}\right)\,\,=\,\,\int_{\epsilon-i\infty}^{\epsilon+i\infty}\frac{d\gamma_{1}}{2\pi i}\int_{\epsilon-i\infty}^{\epsilon+i\infty}\frac{d\gamma_{2}}{2\pi i}\,e^{\gamma_{1}\xi_{1}\,+\,\gamma_{2}\xi_{2}}\Bigg\{\underbrace{\tilde{\rho}_{{\rm in}}^{(2)}\left(\gamma_{1},\gamma_{2}\right)e^{\bar{\alpha}_{S}\left(\chi\left(\gamma_{1}\right)\,+\,\chi\left(\gamma_{2}\right)\right)\,Y}}_{Fig.~(\ref{2cont})-a}\label{SOL}\\
 &  & \,\,+\,\,\underbrace{h\left(\gamma_{1},\gamma_{2}\right)\,\tilde{\rho}_{{\rm in}}^{(1)}\left(\gamma_{1}+\gamma_{2}\right)\,\int_{0}^{Y}dy'e^{\bar{\alpha}_{S}\left(\chi\left(\gamma_{1}\right)+\chi\left(\gamma_{2}\right)\right)\,\left(Y-y'\right)\,\,+\,\,\bar{\alpha}_{S}\chi\left(\gamma_{1}+\gamma_{2}\right)\,y'}}_{Fig.~(\ref{2cont})-b}\Bigg\}\nonumber 
\end{eqnarray}
where \cite{BFKL,KOLEB} $\xi_{1}\,=\,\ln\left(r_{1}^{2}\right)$,
$\xi_{2}\,=\,\ln\left(r_{2}^{2}\right)$ and 
\begin{equation}
\chi\left(\gamma\right)\,\,=\,\,2\psi\left(1\right)\,-\,\psi\left(\gamma\right)\,-\,\psi\left(1-\gamma\right)\,~~~~~~~~~\mbox{where}~~~~~~\psi(z)=\frac{d\,\ln\Gamma(z)}{dz}.\label{CHI}
\end{equation}
The functions$\tilde{\rho}_{{\rm in}}^{(2)}\left(\gamma_{1},\gamma_{2}\right)$
and $\tilde{\rho}_{{\rm in}}^{(1)}\left(\gamma_{1}+\gamma_{2}\right)$
are determined by the initial condition for the two and one parton
shower productions (see Fig.~(\ref{2cont}). The explicit form of
the function $h\left(\gamma_{1},\,\gamma_{2}\right)$ can be found
in Ref.\cite{GOLEPSI}. Different contributions in the solution~(\ref{SOL})
might be visualized as shown in the Figure~\ref{2cont}. 

 \end{document}